%% file: ms.tex
\documentclass{emulateapj}

\begin{document}

\title{EG And: {\it{FUSE}} and {\it{HST}}/STIS Monitoring of an Eclipsing Symbiotic Binary}

\author{Cian Crowley and Brian R. Espey\altaffilmark{1}}
\affil{School of Physics, Trinity College Dublin, Dublin 2, Ireland}
\email{cian.crowley@tcd.ie}
\email{brian.espey@tcd.ie}
\and
\author{Stephan R. McCandliss}
\affil{Department of Physics and Astronomy, Johns Hopkins University, Baltimore, MD 21218, USA}
\email{stephan@pha.jhu.edu}

\altaffiltext{1}{School of Cosmic Physics, Dublin Institute for Advanced Studies, Dublin 2, Ireland}

%\maketitle 

\begin{abstract}

We present  highlights and an overview of 20 {\em FUSE} and {\em HST}/STIS observations of the bright symbiotic binary \object{EG And}. The main motivation behind this work is to obtain spatially-resolved information on an evolved giant star in order to understand the mass-loss processes at work in these objects. The system consists of a low-luminosity white dwarf and a mass-losing, non-dusty M2 giant. The ultraviolet observations follow the white dwarf continuum through periodic gradual occultations by the wind and chromosphere of the giant, providing a unique diagnosis of the circumstellar gas in absorption.  Unocculted spectra display high ionization features, such as the \ion{O}{6} resonance doublet which is present as a variable ({\em hourly} time-scales), broad wind profile, which diagnose the hot gas close to the dwarf component.  Spectra observed at  stages of partial occultation display a host of low-ionization, narrow, absorption lines, with transitions observed from lower energy levels up to $\sim$5 eV above ground. This absorption is due to chromospheric/wind material, with most lines due to transitions of \ion{Si}{2}, \ion{P}{2}, \ion{N}{1}, \ion{Fe}{2} and \ion{Ni}{2}, as well as heavily damped \ion{H}{1} Lyman series features. No molecular features are observed in the wind acceleration region despite the sensitivity of {\em FUSE} to H$_2$. From analysis of the ultraviolet dataset, as well as optical data, we find that the  dwarf radiation does not dominate the wind acceleration region of the giant, and that observed thermal and dynamic wind properties are most likely representative of isolated red giants.

\end{abstract}

\section{Introduction}

For the majority of red giant stars the basic mass-loss processes at work are unknown. Indeed, for stars of spectral types between K0 III and M5-M6 III, much remains unknown about the regions above the visible photosphere and the transportation of processed material to the interstellar medium (ISM). For those stars approaching the tip of the asymptotic giant branch (AGB) the stellar winds are thought to be driven by a combination of stellar pulsations which levitate matter out to regions conducive to dust formation (i.e.\ temperatures below $\sim$10$^3$ K) and subsequent radiation pressure on dust. This combination of pulsations and radiative pressure  can account for the large mass loss rates observed in heavily evolved giants. However, those stars on the first ascent of the red giant branch do not possess significant dusty circumstellar shells and their pulsations are also orders of magnitude weaker. The observations of massive winds from these objects therefore poses a major problem. There has been relatively little progress in our understanding of the physical processes at work in the regions above the photospheres of these giants and current theories are in need of more observational constraints. We have obtained a series of ultraviolet observations of a high-inclination symbiotic star with a view towards determining the thermal and dynamic conditions from the chromosphere through the wind acceleration region of an M2 giant. The primary aim of these observations is to utilize the diagnostic power of the ultraviolet spectral region to examine the circumstellar conditions along a number of lines of sight, providing constraints on the possible processes at work in these regions. 

\subsection{Symbiotic Stars}

Originally identified and classified by their distinctive composite optical spectra, symbiotic stars display optical spectral features associated with both a cool giant star and an ionized nebula. They are now known to be binary systems containing an evolved giant star and a hot white dwarf or subdwarf, with orbital periods of non-dusty  systems typically $\sim$1-3 years. The nebular line emission is attributed to  high energy  photons from the dwarf which ionize a portion of the dense giant wind. %Since their discovery in the 1920s symbiotics have been studied extensively, in particular with the purpose of further understanding stellar evolution. 
For a general overview of symbiotic stars and their features see \citet{1986syst.book.....K} and \citet{1988syph.book.....M}. %The unique configuration of a small, hot continuum source in orbit around an extended mass-losing giant provides us with a means of studying the winds and circumstellar environment of evolved stars. The use of these systems to probe giant winds holds many advantages over other classes of binary stars:

Those  systems which are viewed close to edge-on provide an opportunity to obtain spatially resolved information on the giant's extended atmosphere and wind. The presence of the dwarf star, combined with knowledge of the orbital parameters, make it possible to use the secondary as an orbiting ultraviolet-bright backlight. Observations taken at well chosen orbital phases make it possible to study the circumstellar material in absorption along differing lines of sight, thus providing tomographic information. 

For isolated stars, generally the only wind parameters that can be reliably determined are disc-averaged global properties such as the mass loss-rate and the terminal wind velocity. The role of eclipsing binaries in providing localized information on circumstellar conditions has long been recognized and {\it{IUE}} ({\it{International Ultraviolet Explorer}}) studies of $\zeta$ Aurigae and VV Cepeid systems have been particularly useful in examining the chromospheres and winds of giants and supergiants \citep[e.g.][]{1987IAUS..122..307R,1990iuea.rept...65B}. 

We believe, however, that the study of symbiotic systems holds a number of advantages over other binary systems.  Unlike many other binaries, the two components in symbiotic systems have entirely different spectral characteristics and can be easily disentangled. The giant continuum does not contribute in the far ultraviolet and the high-velocity, high-ionization material close to the dwarf is easily distinguished from the low-velocity, low-ionization giant wind features. Also, the small diameter of the dwarf relative to the giant ($\sim$ 0.0002\%) provides a very narrow, ``pencil beam'' view through the outer layers of the giant. In addition, many non-dusty symbiotics have been well studied across  all wavelengths and the orbital elements and geometrical parameters of a large number are well known. Finally, the periods of these objects are compatible with   space-based observation scheduling requirements permitting a well sampled dataset over the binary orbital period.

Using {\it FUSE} ({\it{Far Ultraviolet Spectroscopic Explorer}}) and STIS (Space Telescope Imaging Spectrograph) data it is possible to obtain high enough signal-to-noise and spectral resolution to resolve the narrow phase-dependent wind features in absorption, which was not possible with {\it{IUE}}.

\subsection{EG And}

The quiescent symbiotic star EG And (HD 4174) is well documented in the literature \citep[e.g.][]{1980ApJ...237..831S,1984ApJ...281L..75S,1985ApJ...295..620O,1991A&A...245..531S,1992A&A...260..156V,1993A&A...274L..21V,1993A&A...273..425M,1995MNRAS.272..189T,1997MNRAS.291...54W} and is considered a prototype for stable, non-dusty symbiotic stars. Indeed it was observed extensively with {\it{IUE}}  where the ultraviolet eclipse effect was used to study the dimensions and wind of the giant star  \citep{1991A&A...249..173V,1992A&A...260..156V}. The object is one of the closest and brightest symbiotic systems and consists of a hot, low luminosity  white dwarf \citep{1991A&A...248..458M} with an M2 \citep{1987AJ.....93..938K,2004AJ....128.2981K} giant primary  which is on the first ascent of the red giant branch. Based on a metallicity analysis and its galactic location, the system belongs to the old disc population \citep[as also found  by][]{1981Obs...101..172W}. The optical spectrum of the giant is very similar to that of isolated   standards of similar spectral type (see \S~\ref{section6}) and far-infrared data from {\it{IRAS}} show fluxes very similar to normal isolated red giants \citep{1986AJ.....92.1118K}. Due to the low-luminosity of the dwarf, the giant's atmosphere is not greatly affected by the presence of the ionizing companion.  In fact, radio measurements confirm that the ionized region around the dwarf is relatively small and does not dominate the cool wind \citep{1990ApJ...349..313S,2003ASPC..303..343S}. 

EG And has never been observed to undergo outburst. Periodic ultraviolet variations are attributed to the ultraviolet source being occulted by the atmosphere of the primary component \citep{1991A&A...249..173V} and has been observed extensively over several orbital epochs by {\it{IUE}}.   An accurate orbit for the system was first determined by \citet{1993A&A...273..425M} and further refined by \citet{2000AJ....119.1375F} based on  velocity measurements of red giant photospheric lines. The high systemic radial velocity (-95 km s$^{-1}$ relative to heliocentric)  permits the wind absorption features to be easily distinguished from interstellar absorption features ($\sim - 30$ km s$^{-1}$ relative to heliocentric), thereby removing the complications that arise when systemic features are merged with interstellar features. Reviewing the attributes of the system, most especially noting the low interstellar extinction, the absence of  circumstellar dust  \citep{1986AJ.....92.1118K,1994AJ....108.1112V}, the low dwarf luminosity, and the proximity of the binary, it is apparent that EG And is an ideal candidate for an ultraviolet wind analysis. System parameters are detailed in Table~\ref{tbl-1}.

In this paper we present an overview of a series of {\it{FUSE}} and STIS observations of EG And, whilst discussing some highlights in more detail. In \S~\ref{section2}  the data reduction and the orbital timing of our observations are outlined. In \S~\ref{section3} we outline the general behavior the ultraviolet spectra over the three orbital epochs and in \S~\ref{section4} we describe the features of the uneclipsed spectra. Those spectra which are partially absorbed by the giant wind and which are used to diagnose the wind conditions, are described in \S~\ref{section5}. Finally, in \S~\ref{section6} we discuss the significance of the spectral variations, the effect of the dwarf on the giant wind and how an analysis of the data can provide improvements in our understanding of the circumstellar environment and winds of evolved giants.
%\clearpage
\input{tab1}
%\clearpage

\section{Observations and Data Reduction} \label{section2}

\begin{figure}
%%\figurenum{1}
\epsscale{1.}
\plotone{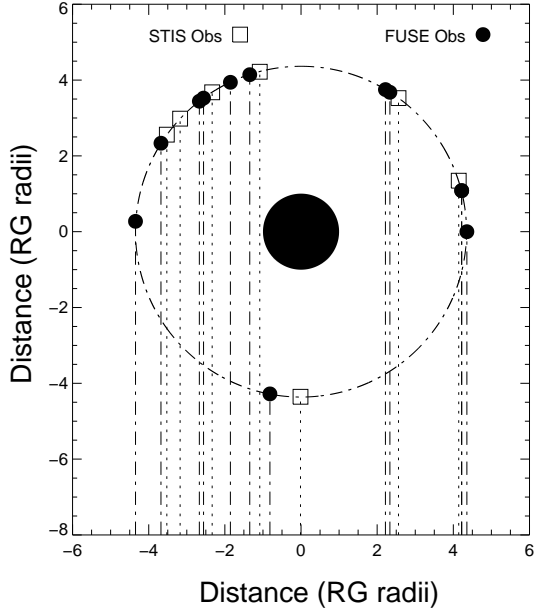}
\figcaption{View of EG And perpendicular to the orbital plane. Circular and square points correspond to the dwarf position for the {\it{FUSE}} and {\it{HST}} observations respectively. Observer's view is from the bottom and the scale is in units of red giant radii. See also Table~\ref{tbl-2}. \label{fig1}}
\end{figure}

The combined {\it{FUSE}} and STIS echelle wavelength coverage (905\AA\ - 3100\AA) provides access to the transitions of many important atomic and molecular species, including  low-ionization species expected to exist in the giant wind as well as those diagnosing the hotter gas associated with the dwarf. 
The ultraviolet observing program was designed to cover orbital phases both in and out of occultation, as well as to test the stability of the system and the repeatability of observations over orbital time-scales. The timing of the observations was designed to include unabsorbed phases (i.e.\ inferior conjunction and quadrature), as well as phases close to ultraviolet minimum, and also a series of ingress and egress observations with intermediate strength absorption.
See Figure~\ref{fig1} for a schematic diagram showing the positions of the dwarf star at the time of the ultraviolet observations.

%Unocculted spectra provide a template for comparison with the absorbed spectra, as well as providing diagnostics of the hot component and the material close to it. This is important as it is not necessary to compare spectra with a different unabsorbed star or perform ab initio modelling of continuum and emission lines or, indeed, ISM components prior to deducing the giant absorption component. 
Comparison of the unabsorbed with the absorbed data allows the diagnosis of the absorbing material; i.e.\ the giant wind. Analysis is simplified by using the unabsorbed spectra as they enable the study of the {\it{ratios}} of spectra to analyze the variations, removing the need to model the unabsorbed continuum, emission and interstellar absorption. % 
Twenty ultraviolet observations of the system were obtained between January 2000 and December 2003. Phase $\phi$ =0.0 is defined as the point at where the dwarf is at superior conjunction, and is completely occulted by the giant and its wind, i.e.\ ultraviolet minimum. Hereafter we reference orbital phases, $\phi$,  against the \citet{2000AJ....119.1375F} ephemeris since this work makes use of all available reliable orbital data measurements and covers the longest baseline. This also enables the orbital epoch  to be identified from the phase information. In the cases where reduced phases are referred to we use $\tilde{\phi}$. See Table~\ref{tbl-2} for observed phases.

%\clearpage
%\clearpage

\subsection{FUSE data}

We have obtained thirteen {\it{FUSE}} observations of EG And (see Table~\ref{tbl-2}). All data were acquired with the target in the large aperture (LWRS; 30\arcsec x30\arcsec) in TTAG photon collecting mode over a time period of 3 {\it{FUSE}} observing cycles. The {\it{FUSE}}  satellite covers the wavelength wavelength region 905-1187\AA\ with a nominal spectral resolution of $\Delta v$ $\sim$ 20  km s$^{-1}$ and has been described by \citet{2000ApJ...538L...1M} and \citet{2000ApJ...538L...7S}. The data were reduced and calibrated with CALFUSE version  3.0.8 \citep{2003ASPC..295..241D}. The effects of spacecraft motion are corrected for with CALFUSE, which then places the data on a heliocentric velocity scale. Interstellar absorption lines in the 8 different channels were compared to interstellar features in the STIS data in order to correct for small wavelength offsets due to target drift in the {\it{FUSE}} aperture. For most observations the data from each channel were shifted and  co-added to increase the signal to noise ratio ($S/N$). However, for those observations where the target was known to drift completely out of the aperture in some channels, the data from each channel were analyzed separately. Exposure times were calculated in order to reach continuum $S/N$ ratios of at least 15 per resolution element for the co-added spectra at 1050\AA.

The thirteen {\em FUSE} spectra  provide good orbital phase coverage, producing spectra displaying differing degrees of wind obscuration. The dataset also contains three observations of the system at very similar phases over 3 separate epochs. These spectra in particular can be used to determine the stability of the system and whether observations taken over different epochs can be compared.  %In addition, three partially absorbed spectra were taken over a time-period of four days, making it possible to study the small-scale distribution of the giant wind.

\subsection{STIS data}

The seven {\it{HST}} ({\it{Hubble Space Telescope}}) observations (see Table~\ref{tbl-2}) were carried out with the medium resolution echelle gratings (E140M and E230M) of the Space Telescope Imaging Spectrograph (STIS) through the 0.2\arcsec x0.06\arcsec\ aperture at the 1425\AA, 1978\AA\ and 2707\AA\ central wavelength settings. This resulted in a  resolving power of $R \sim 30,000-45,000$ ($\Delta v$ $\sim 6-10$ km s$^{-1}$) over the wavelength range $\sim$1150-3100\AA. The observations were designed to provide sufficient exposure time  to achieve at least a $S/N$ ratio of 15 in the continuum at the central wavelength of the E140M setting. This typically resulted in a total of 2 orbits worth of exposure time for all three settings for the unabsorbed phases and a total of 3 orbits for the absorbed phases. The data were reduced within the IRAF environment using the standard Space Telescope Science Institute (STScI) reduction package (CALSTIS).

The STIS observations of EG And cover two  phases which are not affected by the giant wind, one at almost total occultation, and four at partially absorbed phases. The first STIS observation which was taken at phase $\phi$=3.80 was taken only five days after one of the   {\em FUSE} observations. These near-contemporaneous observations provide the opportunity to test the variability of features on short time-scales, as well as  making it possible to compare diagnostically important line-profiles  over the combined wavelength region (i.e.\ resonance lines  of \ion{O}{6}, \ion{N}{5} and \ion{C}{4}).

%\clearpage
\input{tab2}

%\clearpage
\subsection{Optical Data}

Optical echelle spectra of EG And were acquired using the 3.5 m telescope at Apache Point Observatory (APO). Data were obtained on July 31 1999 (corresponding to an orbital phase of $\phi=$1.47), having  a spectral resolution of $R\sim$40,000. Using a prism as a cross-disperser, the APO echelle covers all wavelengths from 3500 to 10400 \AA\ spread over 100 orders. On the same night a comparison spectrum was obtained of the high velocity M2 III spectral standard HD 148349 using the same instrumental setup. The telescope and instrument are fully described by \citet{1995AAS...186.4404Y}.

\section{Spectral Variations: The Orbital Modulation} \label{section3}

The ultraviolet continuum and emission lines are both modulated by the periodic occultation of the hot material by the red giant and its extended atmosphere. In those spectra not affected  by wind absorption the ultraviolet region is dominated by the continuum of the dwarf. Superimposed on this continuum are emission and absorption features from high-ionization species and narrow interstellar absorption lines. The continuum is observed  to rise towards short wavelengths into the {\it{FUSE}} range before becoming severely attenuated by interstellar absorption approaching the Lyman limit. The narrow interstellar absorption lines originate from species such as \ion{H}{1}, H$_2$, \ion{C}{2}, \ion{N}{1}, \ion{N}{2}, \ion{O}{1}, \ion{Si}{2}, \ion{P}{2}, \ion{Ar}{1} and \ion{Fe}{2} and are consistent with a warm ($\sim$500 K) absorbing cloud with a radial velocity of $\sim-30$ km s$^{-1}$. In addition, emission and P-Cygni features from species such as \ion{C}{3}, \ion{N}{3}, \ion{O}{4}, \ion{P}{5}, \ion{Si}{4}, and \ion{O}{6} seeming to originate mainly from material in the photoionized portion of the giant wind and/or the hot gas close to the dwarf are all prominent in the {\it{FUSE}} data at unabsorbed phases. In the  wavelength region covered by the STIS echelle, the white dwarf continuum falls off to long wavelengths where the hydrogen recombination continuum also contributes. Permitted and semi-forbidden emission lines from ions such as \ion{He}{2}, \ion{C}{2}, \ion{C}{3}, \ion{C}{4}, \ion{N}{3}, \ion{N}{4}, \ion{N}{5}, \ion{O}{1}, \ion{O}{2}, \ion{O}{3}, \ion{O}{4}, \ion{Mg}{2}, \ion{Si}{4} and \ion{Fe}{2} are all prominent.

%\clearpage

\begin{figure*}[!t]
%\figurenum{2}
\epsscale{1.}
\plotone{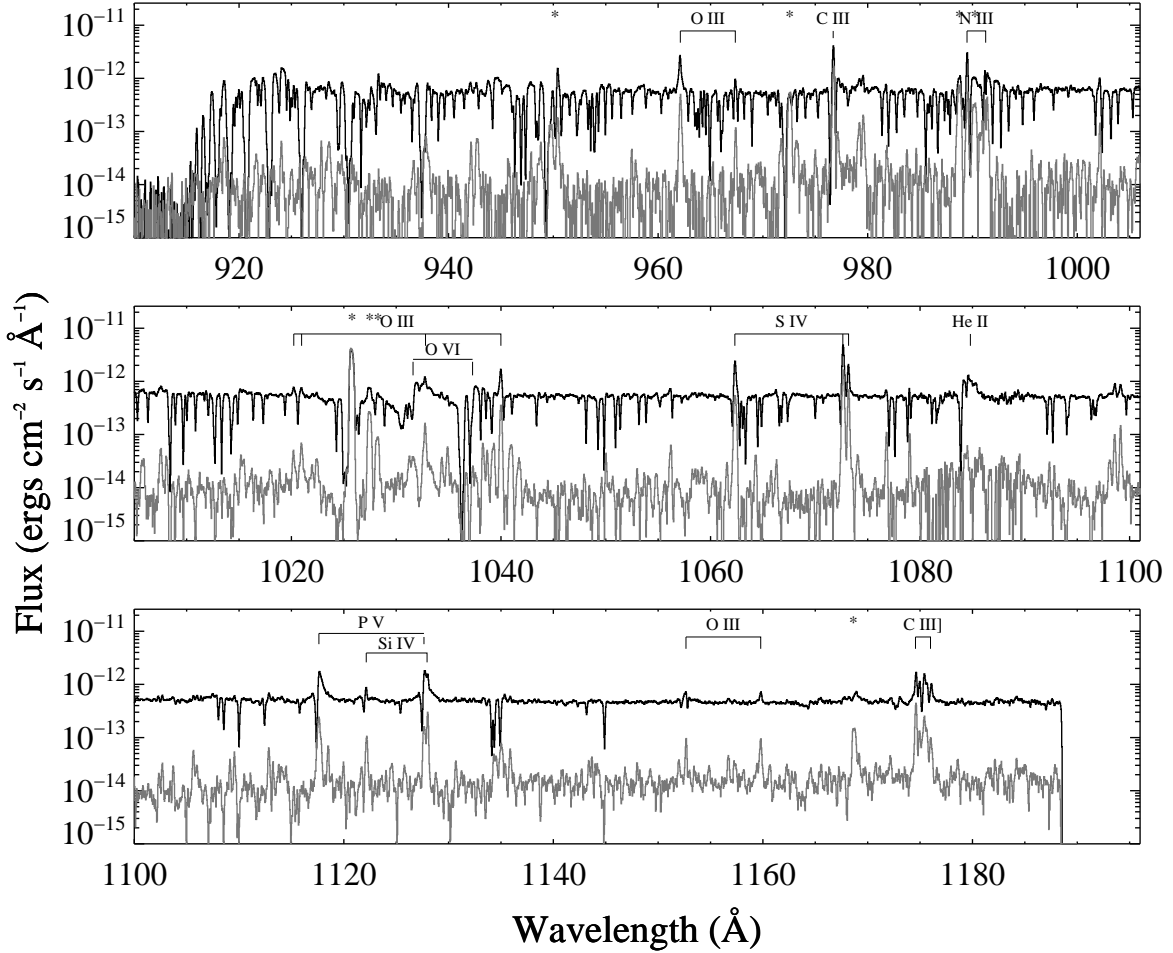}
\caption{{\it{FUSE}} spectra of EG And at quadrature when $\phi$=3.79 (black) and close to totality (gray) when $\phi$=3.05. The y-axis is plotted on a log scale to facilitate display of the strong emission lines and each spectrum is smoothed by the same amount for clarity. All narrow absorption features in the unocculted spectrum are of interstellar origin. \label{fig2}}
\end{figure*}

\begin{figure*}[!t]
%\figurenum{3}
\epsscale{1.}
\plotone{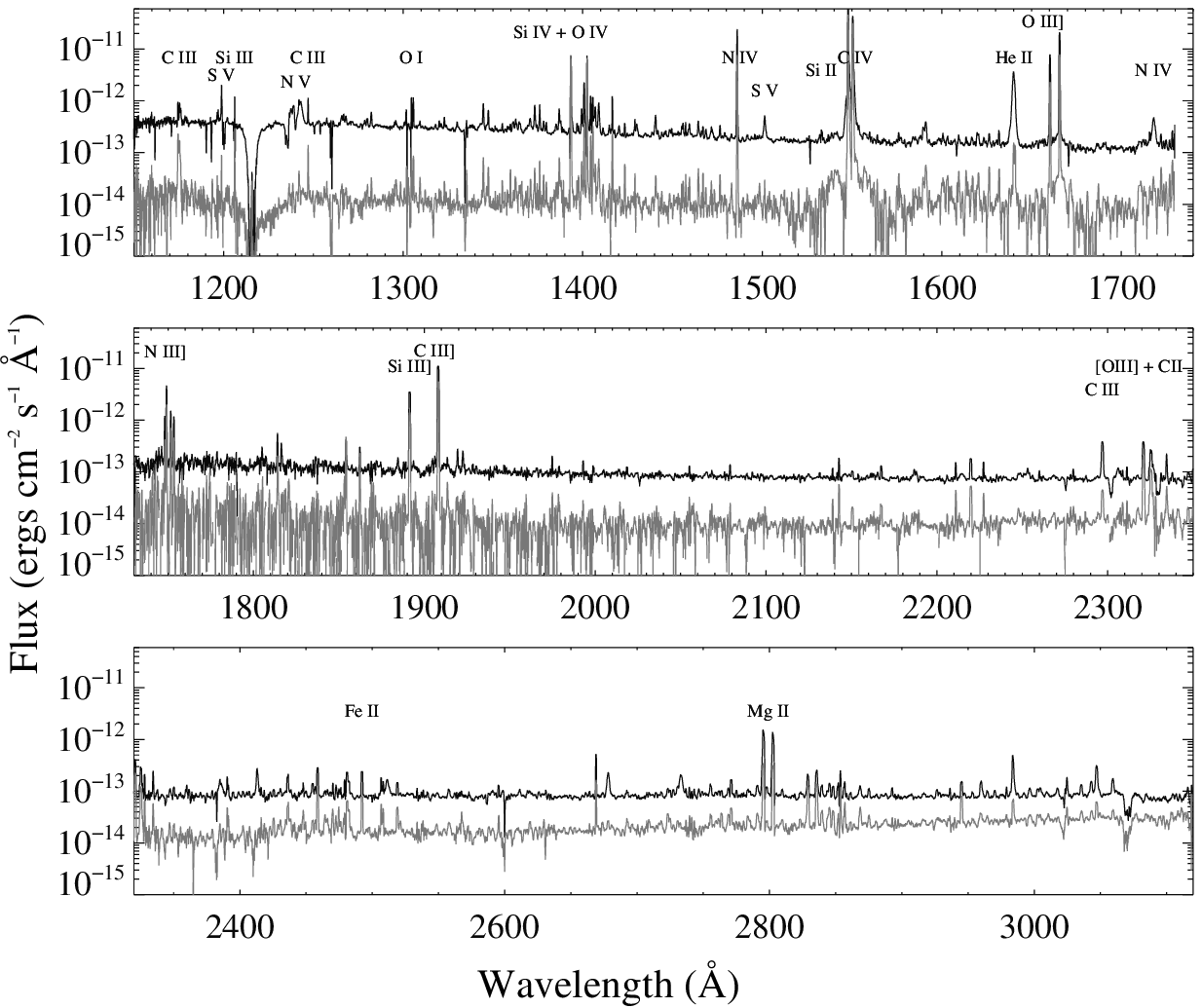}
\caption{STIS spectra of EG And at quadrature (black; $\phi$=3.80) and near ultraviolet minimum (gray; $\phi$=4.04). The y-axis is plotted on a log scale to facilitate display of the strong emission lines. Note the continuum of the giant rising towards red wavelengths in the occulted spectrum.  Data for each spectrum is smoothed by the same amount for clarity. \label{fig3}}
\end{figure*}
%\clearpage
All of these features are  affected by the occultation, during which the broad components of the high ionization lines completely disappear. During occulted phases ($\tilde{\phi} \leq 0.16$) the continuum shape is defined by the damped wings of the hydrogen Lyman series transitions as the dwarf is obscured by large amounts of neutral material in the giant's wind. The large increases observed in the hydrogen column densities are accompanied by the appearance of a host of narrow absorption lines from neutral or lowly ionized species. The strength of these features vary in tandem with the  strength of the \ion{H}{1} lines.

Presented in Figures~\ref{fig2} and \ref{fig3} are plots of quadrature and heavily absorbed spectra taken with {\it{FUSE}} and STIS, respectively. The fluxes are plotted on a log scale to facilitate display of the strong emission lines as well as absorption features. Note, especially, the strong modulation of the ultraviolet continuum and the high ionization lines (i.e.\ \ion{He}{2} 1640 \AA
), whereas lines of species of lower ionization (i.e.\ \ion{Mg}{2} 2800 \AA) are less affected by the `eclipse'.  At longer wavelengths the contribution of the dwarf to the continuum decreases and the dominant contributors are the nebular recombination continuum and the continuum of the red giant itself, explaining why the continuum is less affected by the occultation at longer wavelengths in the STIS data.

\begin{figure*}[h]
%\figurenum{4}
\epsscale{1.}
\plotone{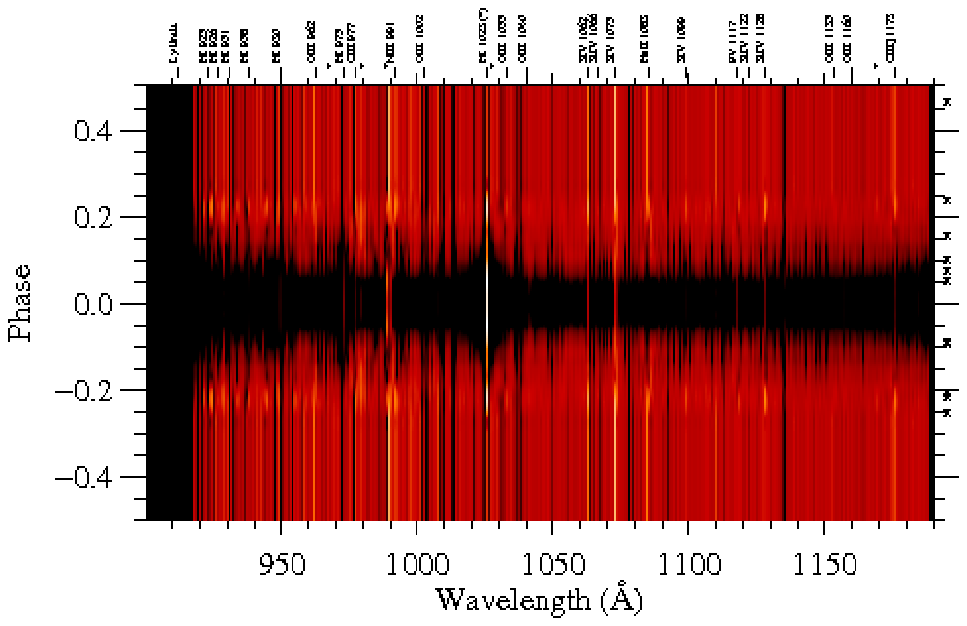}
\caption{ A graphical representation of the spectral variations in the {\it FUSE} EG And data. The flux levels are represented by  color intensities and are displayed on a log scale. The data are phase-wrapped and the flux intensities are linearly interpolated through phases where no data is present.  Note that the `eclipse' effect is much larger at wavelengths close to the hydrogen lines. Spectral artifacts are marked with an asterisk and the phases at which observations took place are marked with a cross to the right of the plot. \label{fig4}}
\end{figure*}

\begin{figure*}[!th]
%\figurenum{5}
\epsscale{1.}
\plotone{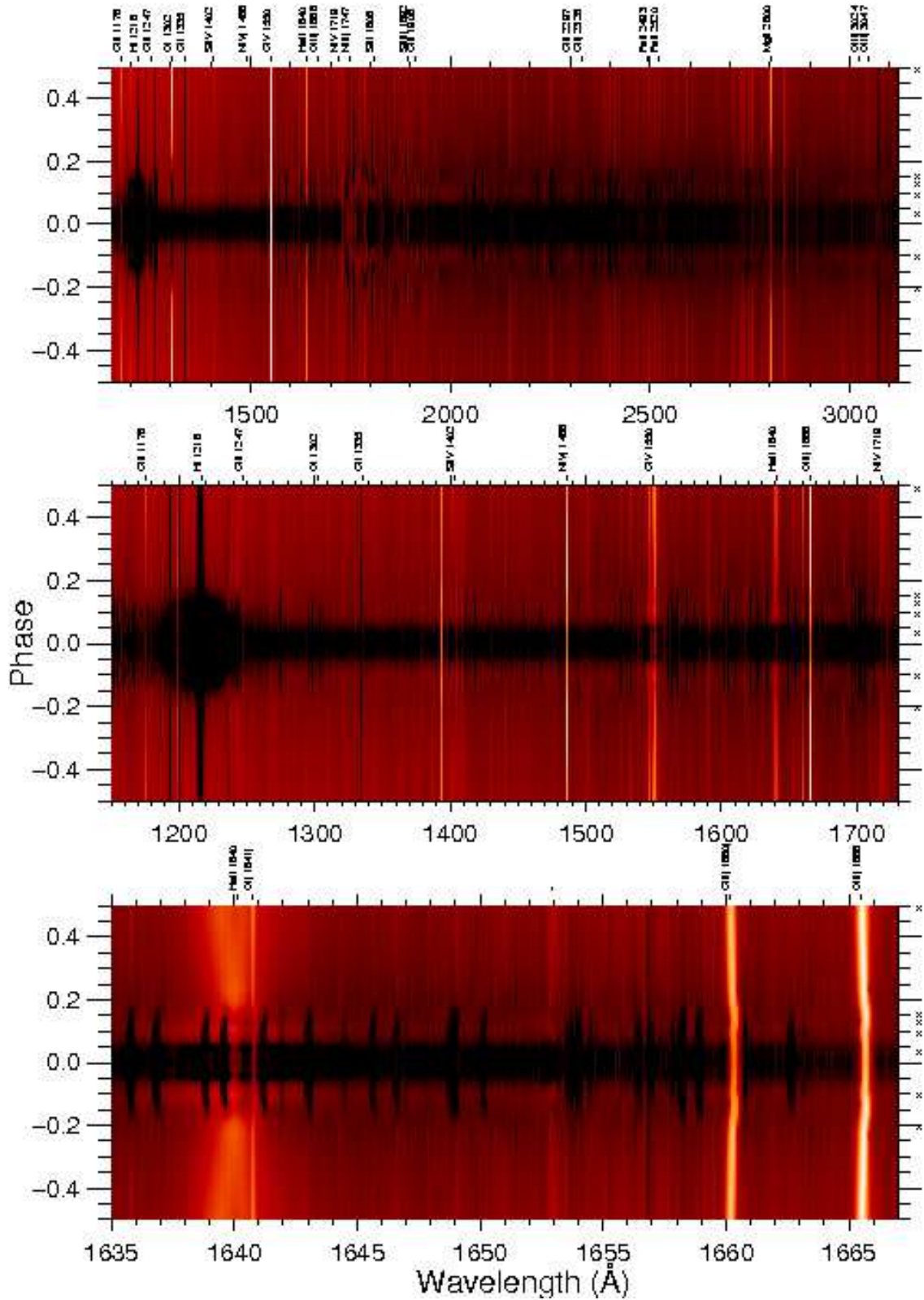}
\caption{ A graphical representation of the STIS ultraviolet data produced using the same technique as was used to generate Figure~\ref{fig4}. The top panel shows the full ultraviolet STIS dataset. The middle panel and lower panels display sections of the same data on a larger scale. Note the differing effect of the occultation on the broad and narrow emission features on the lower panel. Spectral artifacts are marked with an asterisk and the phases at which observations took place are marked with a cross to the right of the plot.\label{fig5}}
\end{figure*}

\begin{figure}[!t]
%\figurenum{6}
\epsscale{1.}
\plotone{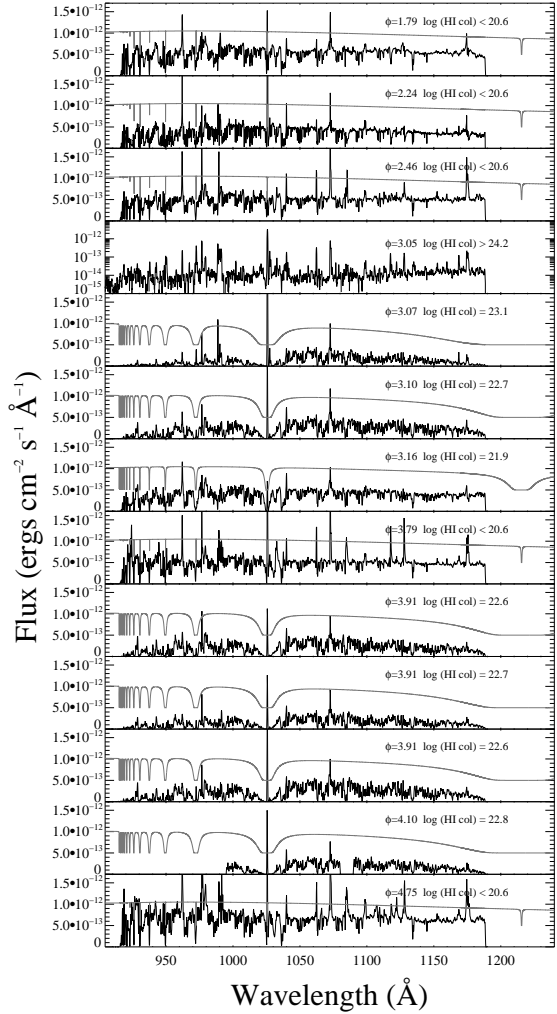}
\caption{Plots of the  complete {\it FUSE} dataset for EG And (time-series from top to bottom). Overplotted  is the continuum of the template unabsorbed spectrum (generated by fitting the continuum of the $\phi$=3.79 spectrum) attenuated by \ion{H}{1} absorption models of differing column densities (labeled on plot).  The data are binned and the models are offset for clarity. \label{fig6}}
\end{figure}

\begin{figure}[!ht]
%\figurenum{7}
\epsscale{1.}
\plotone{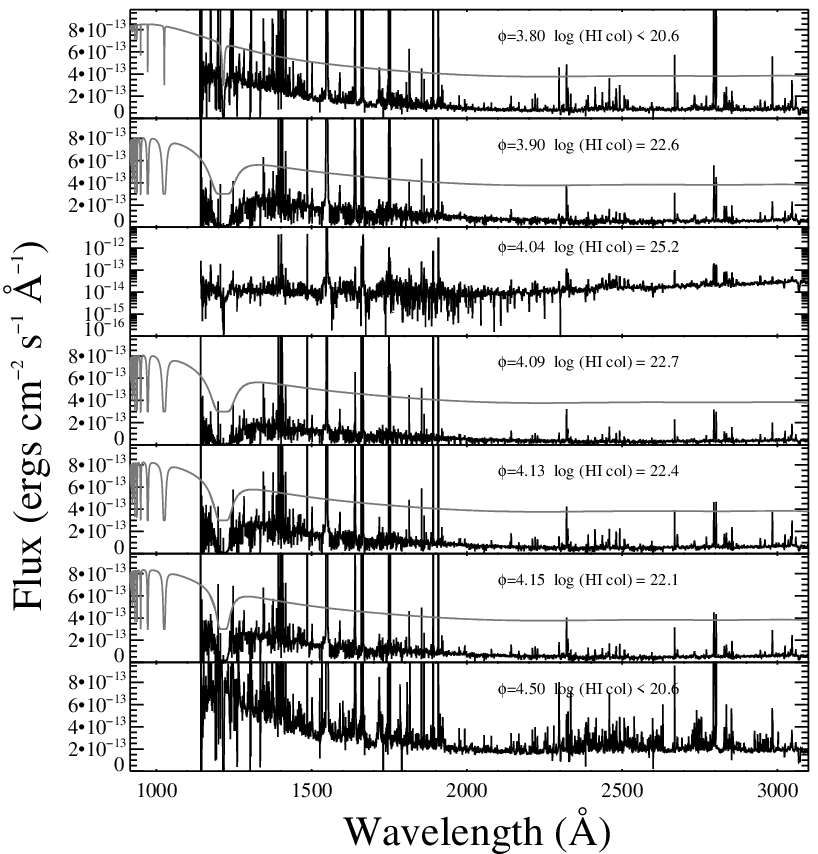}
\caption{Plots of the  complete STIS ultraviolet dataset for EG And, similar to Figure~\ref{fig6}. The template unabsorbed spectrum is derived from a fit to the $\phi$=3.80 spectrum. The data are binned and the models are offset for clarity. \label{fig7}}
\end{figure}

In order to provide an overview of the complete ultraviolet dataset, graphical representations of the variations in the {\em FUSE} and STIS data are presented in Figures~\ref{fig4} and \ref{fig5} respectively. The top panel of Figure~\ref{fig5} shows the full STIS echelle dataset, the middle panel shows only the E140M data, while the lower panel shows the region of the E140M spectrum close to the \ion{He}{2} 1640 \AA\ feature. The flux levels are represented by color intensities and are displayed on a log scale. The data are phase-wrapped and the flux levels are linearly interpolated through those orbital phases where no data is present. The dramatic effects of the orbital modulation on the ultraviolet data are very apparent in these diagrams. The attenuation of the continuum is observed to continue from reduced phase $\tilde{\phi}=$0.00 out to $\tilde{phi}\sim$0.16, with the attenuation being stronger close to the \ion{H}{1} Lyman transitions. The effect of the occultation on various emission lines is illustrated in the lower panel of Figure~\ref{fig5}. The broad \ion{He}{2} 1640 emission feature  originates in material close to the dwarf component, hence the occultation of this feature is almost complete. In contrast, the \ion{O}{3}] 1660, 1666 doublet is much less affected by the occultation of the hot component. This feature originates in an ionized part of the outer wind of the giant which is much more geometrically extended than the \ion{He}{2} emitting zone, and therefore much less affected. The appearance of the narrow wind absorption features on the continuum during partially absorbed phases is also illustrated in this diagram. The majority of the variable absorption lines in this wavelength region originate from  excited levels of the Fe$^{+}$ ion. We also show time-series plots of all of the  {\em FUSE} and STIS spectra are presented in Figures~\ref{fig6} and \ref{fig7}. Also plotted are the \ion{H}{1} models (gray) which are multiplied by fits to the continua of the reference quadrature data.

The general behavior of the ultraviolet spectral variations can be explained in terms of the gradual occultation of both the white dwarf and most of the the photoionized region of the giant wind by the extended atmosphere of the giant itself. However, there are also variations that appear unrelated to orbital phase and this appears to be associated with material close to the dwarf.

%\clearpage

\section{Spectral Features associated with the Dwarf} \label{section4}

It was originally anticipated that the wavelength coverage of {\it{FUSE}}, combined with the comprehensive orbital coverage and the high  quality of the observations, would permit the identification of white dwarf photospheric features and establish the system as a double-lined spectroscopic binary. However, the only direct observation of the hot star is the ultraviolet continuum. The broad components of high-ionization emission lines from species such as \ion{He}{2}, \ion{C}{4}, \ion{N}{4}, \ion{N}{5} and \ion{S}{5} appear to originate in gas close to the dwarf, however wind absorption features, self-absorption and line blanketing preclude reliable radial velocity measurements and no dwarf photospheric features have been identified. However, analysis  of  these broad line profiles places this  material in very close proximity to the dwarf.  We firstly discuss the resonance profiles that appear in P-Cygni form, followed by an overview of the broad lines which have no absorbing component. %See Figures~\ref{fig8} and \ref{fig9} for plots (on a log scale and in velocity space) of a sample emission features in the {\em FUSE} and STIS data respectively.
%%\clearpage
%\begin{figure*}[!th]
%\epsscale{.8}
%\plotone{f8.eps}
%\caption{A selection of emission lines from each  {\em FUSE} spectrum shown on a log scale and in velocity space. The specified f number for each line denotes the flux reduction factor required in order for all lines to be plotted on a common flux scale. Note the changes in scale of the x-axis for each line and that the data are smoothed for clarity. The overplotting of both components of the \ion{O}{6} doublet permit the intrinsic broad \ion{O}{6} profile to be isolated from narrow features associated with the wind or ISM. Note that at more absorbed phases that the short wavelength component is affected by absorption by the wing of \ion{H}{1}  Lyman $\beta$ absorption. \label{fig8}}
%\end{figure*} 

%\begin{figure*}[!th]
%\epsscale{.8}
%\plotone{f9.eps}
%\caption{A selection of emission lines from each  STIS spectrum shown on a log scale and in velocity space, with a similar format to Figure~\ref{fig8}. The overplotting of both components of the \ion{N}{5} doublet permit the intrinsic broad \ion{N}{5} profile to be isolated from narrow features associated with the wind or ISM. Note that at more absorbed phases that the short wavelength component is affected by absorption by the wing of \ion{H}{1}  Lyman $\alpha$ absorption. \label{fig9}}
%\end{figure*} 
%%\clearpage

%The fact that the interstellar H I absorption masks any possible photospheric hydrogen features suggests that the dwarf has a high surface gravity and/or is hydrogen deficient resulting in narrow Lyman absorption profiles.

\subsection{Broad P-Cygni Features}

A number of permitted, high-ionization transitions are present in the form of broad  P-Cygni, or inverse P-Cygni, profiles. In some cases, these profiles are blended with a narrow emission component from the nebular region, while in others they are mutilated by interstellar absorption at the blue end of the profile.
 
\begin{figure}[!t]
%\figurenum{5}
\epsscale{1.}
\plotone{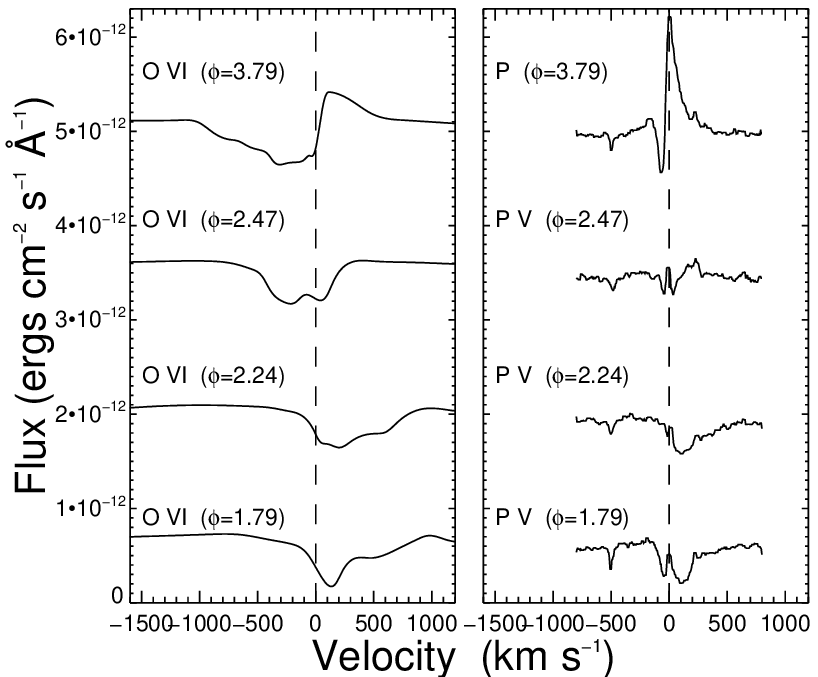}
\caption{The \ion{O}{6} (left panel) and \ion{P}{5} (right panel) resonance line profiles (from unocculted phases) in velocity space in the EG And rest frame. The \ion{O}{6} profiles are reconstructed from clean regions (free of narrow absorption lines) from the each component of the doublet, while the \ion{P}{5} profile is the $\lambda$ 1117 feature. The variations are not related to orbital phase, and the profiles are observed to switch between P-Cygni and inverse P-Cygni form. Although the profiles reach different extents in velocity space, it is apparent that they vary in tandem. These profiles trace  clumpy, dense material located close to the dwarf star. The narrow emission component of the \ion{P}{5} at rest velocity line is a nebular emission component superimposed on the wind profile. \label{fig10}}
\end{figure}

The \ion{O}{6} resonance transitions(1032, 1036 \AA) represent the highest ionization transitions in the spectrum and are present as optically thick wind profiles. They are observed to vary between a P-Cygni form, typical of fast ($\sim$1,000 km s$^{-1}$) expanding winds, and broad red-shifted absorption profiles typically observed in objects with in-falling material present. The variations are phase-independent and trace instabilities in the region deduced to be close to the white dwarf, as will be explained later in this section. See Figure~\ref{fig10} for a plot of the \ion{O}{6} profiles at four unocculted phases as well as  velocity plots of the \ion{P}{5}  1117 \AA\ resonance feature at the same orbital phases. Although P$^{+4}$ has a lower ionization energy than O$^{+5}$, it is apparent that both lines diagnose the same mass-motions, with profiles switching from P-Cygni form to inverse P-Cygni form in tandem.

%\clearpage

%\clearpage

From the {\it{IUE}} observations of EG And, no definitive identification of the \ion{N}{5}  1238, 1242 \AA\ emission features was possible, raising the possibility that one of the component stars has an anomalous abundance \citep{1992PASP..104...87S}. However,it is apparent from our STIS spectra that \ion{N}{5} emission {\em is} present. The profiles are complicated due to self absorption and also because the widths of the features ($\sim$1,000 km s$^{-1}$) are greater than the separation of the doublet. This results in mutilated and overlapping profiles that give the impression of weak emission and an anomalous doublet intensity ratio. The emission disappears during at absorbed phases, which is  expected for gas close to the dwarf.

The \ion{C}{4}  1548, 1550 \AA\ emission feature is the strongest in the ultraviolet spectrum. It is composed of at least two components: a broad base (full width at zero intensity (FWZI) of the convolved doublet $\sim$1400 km s$^{-1}$) which disappears around `eclipse', and a narrower central component which is affected by self absorption. The intra-doublet ratio of the  line core deviates from the optically thin 2:1 ratio to give ratios varying from between 1.5 and 1.8, depending on phase. The  profile is similar to those of the higher ionization lines described above, except that the broad profile has a strong nebular emission line core superimposed on it. The \ion{C}{4} profile and its variations are similar in form to other high-ionization permitted resonance transitions such as \ion{S}{4} and \ion{Si}{4}. A detailed modeling of such resonance line profiles is complicated by the multi-component structure and in many cases self- and interstellar absorption. However, it is readily apparent that there are at least two distinct emitting regions. One in fast-moving gas close to the dwarf and one which is further away from the dwarf, in an ionized region of the outer red giant wind.

\subsection{Broad Emission Lines}

\begin{figure}[!t]
%\figurenum{9}
\epsscale{1.}
\plotone{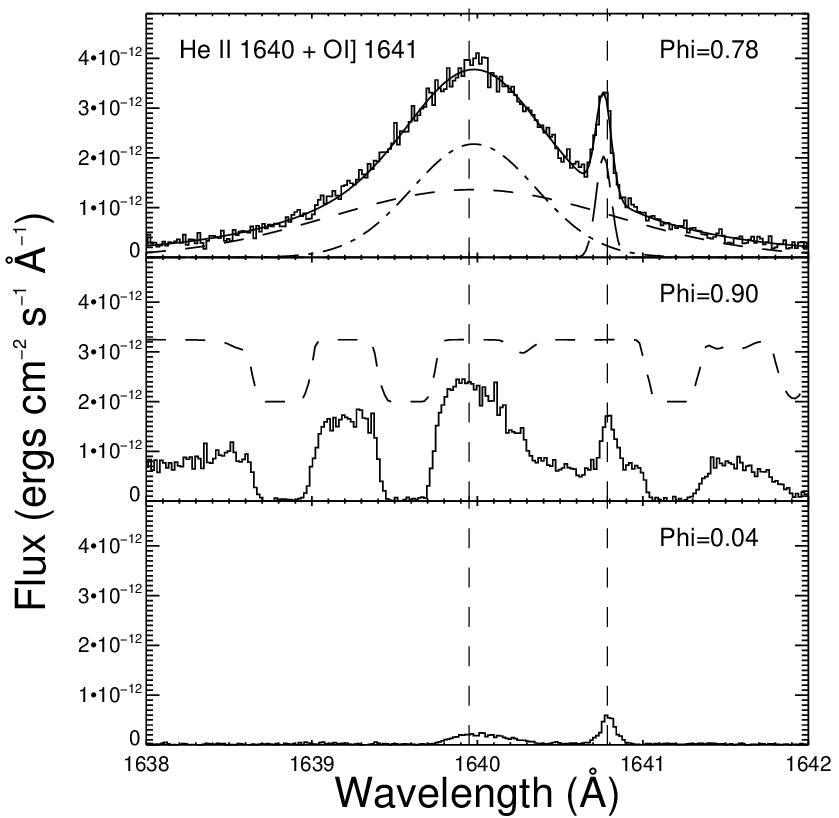}
\caption{The \ion{He}{2} 1640 recombination line at three orbital phases. In the top panel note that the profile is composed of a wide base and a narrower core component. Also note the much narrower \ion{O}{1}] 1641 emission line.  From radial velocity measurements and photoionization modeling (see \S~\ref{section6}) we find that the broad \ion{He}{2} component is located close to the dwarf. In the partially absorbed spectrum displayed in panel 2 the overlying wind absorption from \ion{Fe}{2} is apparent. Overplotted (dashed line) with a vertical offset is an \ion{Fe}{2} absorption model, thermally excited to a chromospheric temperature of 8000 K (which is consistent with our measurements). The bottom panel displays the extent to which the \ion{He}{2} emitting regions have been occulted at $\phi$=4.04. The emitting region is therefore close to the dwarf and is located along the axis between both components. The dashed vertical lines correspond to the rest wavelengths of the lines in the red giant rest frame. \label{fig11} }
\end{figure}

The \ion{He}{2}  1640 \AA\ recombination line  also traces the hotter gas in the system. The emission feature can be decomposed into two Gaussian profiles (see Figure~\ref{fig11}). The broader component is observed to disappear completely during `eclipse', while the emission core is present but greatly reduced at phase $\phi$=4.04. The profile is greatly affected by a number of narrow \ion{Fe}{2} absorption lines in partially absorbed spectra. This explains the asymmetric profiles observed with {\it{IUE}} where self-absorption was suggested in order to explain the profile \citep{1992PASP..104...87S}. (Due to the high energy of the lower level of the transition (40.8 eV), self-absorption would require large amounts of hot material to be passing intermittently  along the line of sight, complicating the model of the system.)

%\clearpage
%\clearpage

The \ion{S}{5} line at 1502 \AA\ (lower level is 15.8 eV above ground) and the \ion{N}{4} line at 1718 \AA\ (lower level is 16.2 eV above ground) appear to originate in a similar emitting region to \ion{He}{2}. They are of similar width (full width at half maximum (FWHM) $\sim$300 km s$^{-1}$), vary in the same way and produce radial velocity measurements that vary out of phase with those features associated with the giant. We find that the amplitudes of the radial velocity shifts for lines of the broad components of these permitted lines of \ion{He}{2}, \ion{S}{5} and \ion{N}{4} are consistent with a dwarf 3 to 4 times less massive than the giant. The lower level of these lines are well above ground, removing complications introduced by self-absorption and, although the profiles are heavily mutilated at absorbed phases, it is noted that the radial velocities of all three broad components behave similarly. 

%Figure~\ref{fig12} shows  a plot of the variation of the radial velocities of the broad component of the   \ion{He}{2} 1640 \AA\ line  and the narrow ($\sim$17 km s$^{-1}$)   \ion{O}{1}] 1641 \AA\ line with orbital phase. The semi-forbidden \ion{O}{1} line shares an upper level with the permitted \ion{O}{1} triplet at $\lambda$ 1302\AA, all of whose transition probabilities are a factor of $\sim$ 10$^{4}$ times higher. Therefore, extremely high optical depths must exist in the line-formation region for an appreciable number of 1641 \AA\ photons to escape from the emitting region via the semi-forbidden transition. It is therefore unsurprising that the measured radial velocities for this line match the predicted radial velocity curve for material associated with the giant atmosphere where densities of neutral material are  high. 

\subsection{Origin of the High Velocity Features}

The observed profiles of the high-ionization lines  can  be understood in terms of the photoionization and accretion of the red giant wind close to the hot component.  The extent of the wind in velocity space depends on the ionization energy of the species; i.e.\ the higher ionization transitions trace higher velocity gas closer to the dwarf. %This is illustrated in Figure~\ref{fig13} where the profiles were reconstructed using splined fits to both components of the doublets in order to exclude self absorption effects and overlying/underlying absorption/emission. 
A simple photoionization argument places the higher ionization material closer to the dwarf, which is where the highest velocities are expected for accreting material (see \S~\ref{section6}). This matches what is observed in the data, and also what is suggested from the radial velocity variations of the unabsorbed broad emission features.

The fact that profiles are observed to switch between P-Cygni and inverse P-Cygni form rules out the existence of a smooth outflow from the hot star. Indeed, for such variations to occur, the material must be reasonably dense and clumpy, suggesting erratic accretion of the cool wind. High densities of this material are also suggested by the absence of broad semi-forbidden line profiles. Unlike some  symbiotic objects such as AG Peg \citep[e.g.,][]{1995A&A...293L..13N}, the broad components in EG And are only observed in the permitted lines. The high ionization semi-forbidden lines exist only as narrow line cores ($\sim$30 km s$^{-1}$), with no hint of a broad component. %Analysis of the \ion{Si}{4} and \ion{O}{4} permitted and resonance multiplets at $\sim$1400\AA\ \citep{2002MNRAS.337..901K} in particular suggest that the semi-forbidden lines are collisionally suppressed.

There is also evidence for an asymmetry in the distribution of hot material around the dwarf.
During the absorbed phases when the broad lines disappear it is apparent that they are modulated asymmetrically around the mid-point of the occultation. The broad wings of lines such as \ion{He}{2}, \ion{C}{4} and \ion{Si}{4} take longer to recover during egress than they do to disappear on ingress. This highlights the asymmetry of the highly ionized region around the dwarf, where the ionizing photons can penetrate further into the less dense material in its wake than into the denser wind material that it is moving into. The asymmetry is thus a line-of-sight effect where we are viewing the wake of the hot component more clearly during ingress phases. %Again, for this effect to be noticeable, the source of the high-velocity emission must be located close to the dwarf, otherwise the total eclipse and orbital asymmetry of these features would not be so pronounced. 

\subsection{Profile Variations Over {\it FUSE} Orbits }

Further evidence for an accretion origin for the broad line profiles  
 comes from the short-timescale variation of the profiles in the {\em FUSE} spectra. When the unabsorbed {\em FUSE} spectra are split into their individual orbital exposures, it emerges that in two of the datasets, the resonance profiles of \ion{O}{6} and \ion{S}{6} are observed to shift dramatically over {\em FUSE} orbital timescales ($\sim$90 minutes) by up to $\sim250$ km s$^{-1}$. The rapid variation of these features is consistent with dense, clumpy material being accreted onto the dwarf component over a very small physical extent. A similar phenomenon is observed in $\zeta$ Aurigae (binary period $\sim$972 days),  where the \ion{C}{4} resonance profiles change over a time-span of $\sim$10 hours (P. Bennett 2005, private communication). These short-timescale variations in the broad line profiles are only present in a small number of the observations. Note that this effect is not observed for any lines of lower ionisation, which are expected to originate further from the dwarf. Indeed moderately ionised lines such as \ion{O}{3} are expected to originate in the extended wind of the giant with a component possibly originating from the heated giant atmosphere \citep[i.e.\ also see][]{1989A&A...208...63M}.

\section{The Giant Chromosphere and Lower Wind} \label{section5}

In contrast to the hot material associated with the dwarf star, the giant's chromosphere and lower wind appears remarkably stable and is relatively unperturbed by the presence of its hot companion. The spectra taken close to ultraviolet minimum provide an uncontaminated view of the giant's chromospheric emission and near-ultraviolet continuum, while the absorbed spectra permit an analysis of the wind distribution and conditions at different impact parameters.

\subsection{Absorbed Spectra}

In all 12 spectra where $\tilde{\phi}\leq$ 0.16, the line-of-sight to the dwarf passes through a significant column of cool wind material, resulting in a host of narrow (FWHM$\sim$7-12 km s$^{-1}$) absorption features superimposed on the dwarf spectrum. As the dwarf becomes further obscured by the wind, the absorption lines grow stronger and the dwarf continuum shape is redefined by attenuation due to the large amounts of neutral hydrogen present in the wind as originally noted by \citet{1991A&A...249..173V}. In both of the spectra with $\tilde{\phi} <$0.06, the continuum is obliterated and what remains is a low level of flux ($\sim$10$^{-14}$ erg s$^{-1}$ cm$^{-2}$ \AA$^{-1}$ ) with nebular and chromospheric emission lines of reduced intensity.

\begin{figure}[!t]
%\figurenum{2}
\epsscale{1.}
\plotone{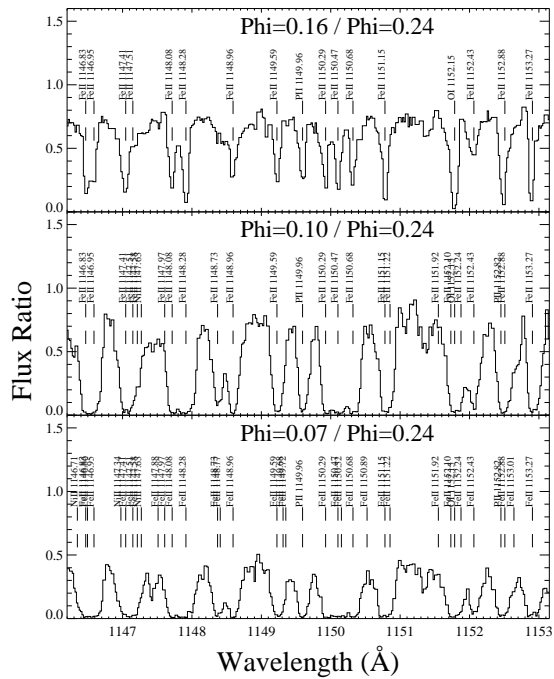}
\caption{Plots of the {\it{ratios}} of an unabsorbed FUSE spectrum to 3 absorbed {\it{FUSE}} spectra at differing degrees of occultation. This plot basically removes the unchanging features (such as interstellar absorption lines) and displays only the spectral variations. The spectra are observed at increasing degree of wind obscuration from top to bottom. Note how the wind absorption lines strengthen and the continuum decreases with respect to the template quadrature spectrum. Most of the absorption features at these wavelengths are due to \ion{Fe}{2}, \ion{P}{2}, \ion{O}{1} and \ion{Ni}{2}.  \label{fig16} }
\end{figure}

Although we have encountered problems in identifying many absorption lines diagnosing the wind, especially within the {\it{FUSE}} bandpass, we have used the compilations of \citet{1995all..book.....K}, \citet{1998yCat..33400300R} and \citet{2003ApJS..149..205M} to identify the vast majority of the absorption features. The wind is predominantly neutral in H, C, N, O and is mainly singly ionized in the heavier common elements such as Mg, Si, S, Ni and Fe. Identified wind absorption features originate from species that include \ion{H}{1}, \ion{C}{1}, \ion{C}{2}, \ion{N}{1}, \ion{N}{2}, \ion{O}{1}, \ion{Mg}{2}, \ion{Al}{2}, \ion{Si}{2}, \ion{P}{2}, \ion{Ar}{1}, \ion{Ti}{2}, \ion{Ti}{3}, \ion{Mn}{2}, \ion{Fe}{2} and \ion{Ni}{2}. The majority of the lines are from singly ionized states of complex iron-group ions such as Mn$^+$, Fe$^+$ and Ni$^+$ and originate from either ground levels, or from lower levels up to $\sim$5 eV above ground. See Figure~\ref{fig16} for plots of the ratios of absorbed FUSE spectra to an unabsorbed {\it{FUSE}} spectrum for 3 differing degrees of occultation. Taking ratios of absorbed to unabsorbed spectra makes use of the unabsorbed spectra to remove the unchanging features (such as interstellar absorption lines) and displays only the variations in emission and/or absorption.

%\clearpage
%\clearpage

\subsubsection{Molecular Lines}
The {\it{FUSE}} wavelength range contains a large number of strong molecular hydrogen transitions from a range of energy levels. Indeed these spectra include a number of H$_2$ lines of interstellar origin. Our observations, therefore, are extremely sensitive to the presence of H$_2$ in the wind acceleration region at a distance  of $\sim$ 2.2 - 3.7 R$_{RG}$. The presence of molecular hydrogen in the wind acceleration region has been postulated as being important for driving the wind \citep{1994AJ....108.1112V}.  No H$_2$ lines (or CO or H$_2$O lines in the STIS data) at the rest velocity of the binary are detected at any phase, setting tight constraints on the molecular content of the wind at these distances above the limb. The upper limit of molecular hydrogen is found to be n(H$_2$)/n(H) $<$ 10$^{-8}$ throughout the wind acceleration region. These results rule out the possibility of molecular opacity (and also, indirectly, dust) contributing to the driving of the wind.

\subsubsection{Wind Structure and Ionization}

By analyzing the absorption lines present in the partially absorbed spectra it is possible to obtain accurate spatially resolved information on the conditions in the wind. In contrast to the complex ionization structure observed in similar eclipsing systems such as the $\zeta$ Aurigae binaries \citep[e.g.][]{1996ApJ...466..979B}, we find that the observed ionization structure along each line of sight is relatively straight-forward. 

\begin{figure}[!t]
%\figurenum{16}
\epsscale{1.}
\plotone{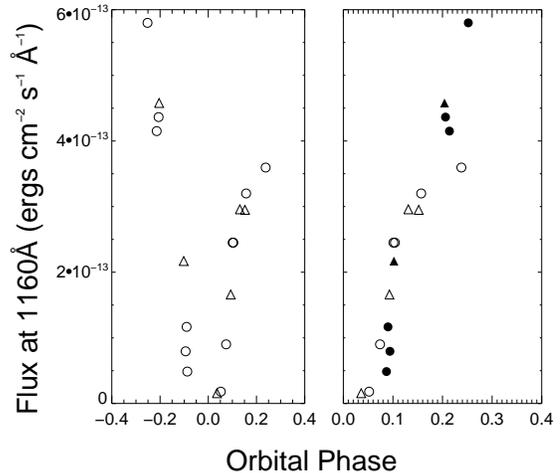}
\caption{Plot of the continuum flux at a line free region at  1160\AA\ against orbital phase phase in the left panel. The right panel shows the same data plotted against the reduced phase. Triangular points represent STIS data and circles represent {\it{FUSE}} data for both plots.  Filled datapoints represent ingress data and open datapoints define egress data for the panel on the right. The observed slight asymmetry between ingress and egress datapoints diagnoses the asymmetry in the distribution around the giant. The variations in fluxes at phases $>$0.20 are due to variations in the continuum of the hot component, and not related to asymmetry in the wind material. The error bars on the flux measurements are are typically less that the size of the datapoints. \label{fig17} }
\end{figure}

Most species exist in only one ionization stage (i.e.\ Fe is present predominently as Fe$^+$) and the ionization is symmetric around `eclipse'. The wavelength coverage provided by {\it{FUSE}} covers the transitions  of many ionization stages for many species, ensuring that any large changes in ionization will be noted (i.e.\ strong transitions from lower levels of Fe$^0$, Fe$^{+}$ and Fe$^{+2}$ are all covered).
% The one exception where changes in ionization {\em are} observed, occurs in the {\it{FUSE}} spectrum taken at $\phi=$0.16, where the Fe absorption is split between Fe$^+$ and Fe$^{+2}$. In this spectrum the linewidth also decreases significantly compared to the other absorbed spectra (i.e.\ from $\sim$12 km s$^{-1}$ to $\sim$7 km s$^{-1}$) and it can also be deduced that a significant amount of hydrogen  in the line of sight is ionized at this phase. This observation probes the line of sight through the wind which is farthest from the giant photosphere, at an impact parameter of $\sim$3.7 R$_{RG}$.
In contrast to the symmetry of the ionization structure around the primary, we observe an asymmetry in the column density of material  that is apparent from both the continuum level variations and is also  from the strengths of the narrow absorption lines. Plotted in Figure~\ref{fig17} are the variations of the continuum flux at a line-free region of the continuum at $\sim$1160\AA. For absorbed spectra the continuum flux at this wavelength is essentially defined by the strength of the damped blue wing of the Lyman $\alpha$ transition of hydrogen. The observed fluxes therefore diagnose an asymmetry in the density of neutral material around `eclipse'. This can be ascribed to a mechanical redistribution of the giant wind by the motion of the dwarf star. From Figure~\ref{fig17} it is apparent from the continuum fluxes at the wind absorbed phases ($\phi<$0.20) that the fluxes on ingress are lower than on egress at the corresponding reduced phase.

%\clearpage
%\clearpage

By modeling the excitation, ionization, line profiles and strengths of the absorption lines it is possible to derive spatially resolved information on the thermal and dynamical conditions throughout the wind acceleration region.
A full analysis and modeling of the wind of the giant will be published in subsequent work.

\subsection{Fully Obscured Spectra}

As previously noted by \citet{1980ApJ...238..929S} and in subsequent papers, the {\it{IUE}} spectrum of EG And during occultation consists of strong narrow emission lines arising from a wide range of excitation energies, with no indication of continuum above the sensitivity limit of {\it{IUE}} ($\sim$1x10$^{-14}$ erg cm$^{-2}$ s$^{-1}$ \AA$^{-1}$). We find that the emission from the broad, high ionization components disappears during occultation and the narrow emission cores ($\sim30-50$ kms$^{-1}$) of the strong emission lines are reduced in strength, typically by a factor of $\sim$3. This residual emission originates from the  photoionized section of the wind between both component stars that is not obscured by the giant. 

There is also a  chromospheric component to the emission lines of the relatively low-ionization and low-excitation lines. This is especially apparent for a number of low-excitation \ion{Fe}{2} and \ion{Ni}{2} emission lines at the long wavelength end of the STIS data. These lines have  upper energy levels that can be collisionally populated at chromospheric temperatures (i.e.\ $\la$10,000K) and, in contrast to the majority of wind features, they posses an emission component at all orbital phases, including those closest to complete occultation. Those transitions that lie within the {\it{FUSE}} and STIS E140M bandpasses (corresponding to $\lambda$ $<$ 1730\AA) must have upper energy levels of at least 7.2 eV (higher if they are not resonance transitions). These energies are too high for the levels to be populated collisionally at chromospheric temperatures. This would explain why the chromospheric emission lines are only observed at the long wavelength end of the STIS ultraviolet data, where the upper levels can be as low as $\sim$ 4 eV above ground. The upper levels can certainly be populated by non-collisional processes, however previous studies of red giant chromospheres have found that \ion{Fe}{2} chromospheric emission is primarily collisionally excited \citep{1991ApJS...77...75J}, in agreement with our observations. %The collisionally excited model for the chromosphere and lower wind is also consistent with the excitation structure derived from our absorption data. 

\subsubsection{Continuum}

The sensitivity limits of the {\it{FUSE}} and STIS detectors offer a considerable improvement on those of {\it{IUE}}. Combining this advantage with the photon-counting nature of the detectors it is possible to observe continuum with flux $<$1x10$^{-14}$ erg cm$^{-2}$ s$^{-1}$ \AA$^{-1}$ for the occulted spectra for both {\it{FUSE}} ($\phi$=3.05) and STIS ($\phi$=4.04). The low-level far-ultraviolet continuum is  due to the scattering of a small proportion of white dwarf photons around the giant atmosphere and into our line of sight. Interpreting the continuum in terms of an incomplete occultation would require the ultraviolet photons to pass through a huge column density of neutral hydrogen, resulting in characteristic absorption surrounding Lyman $\alpha$. The columns involved would be large enough to observed, yet the binned continuum resembles that of the unocculted star, though on a much lower flux level ($\sim$1x10$^{-14}$ erg cm$^{-2}$ s$^{-1}$ \AA$^{-1}$ at 1300\AA). At the long wavelength end of the STIS data the continuum is observed to rise towards red wavelengths (see Figure~\ref{fig3}), which is the beginning of the red giant photospheric continuum which  dominates at optical wavelengths \citep[for further detail see][]{2005A&A...440..995S}.

\section{Discussion} \label{section6}

This set of observations is the most complete set of ultraviolet `eclipse' observations of any symbiotic system. Indeed the orbital sampling and the high resolution and $S/N$ of the data enable us to study a red giant chromosphere and wind in absorption over a range of differing impact parameters from the photosphere. A study of this type can provide observational constraints required in order to understand the mass-loss processes at work in these objects. However, it needs to be clarified by how much the wind is affected by the presence of the ionizing companion, and whether the highly variable behavior of the broad, hot material can be separated from the cool outflow from the giant. %For example, if the wind conditions are found to be  dominated by the ultraviolet radiation field it would be difficult draw general conclusions as regards isolated giants based on this type of analysis. 
The remainder of this section treats the influence of the white dwarf on giant and the material in its inner wind.

\subsection{Effect of the Dwarf on the Giant Atmosphere}

\begin{figure}[!t]
%\figurenum{16}
\epsscale{1.}
\plotone{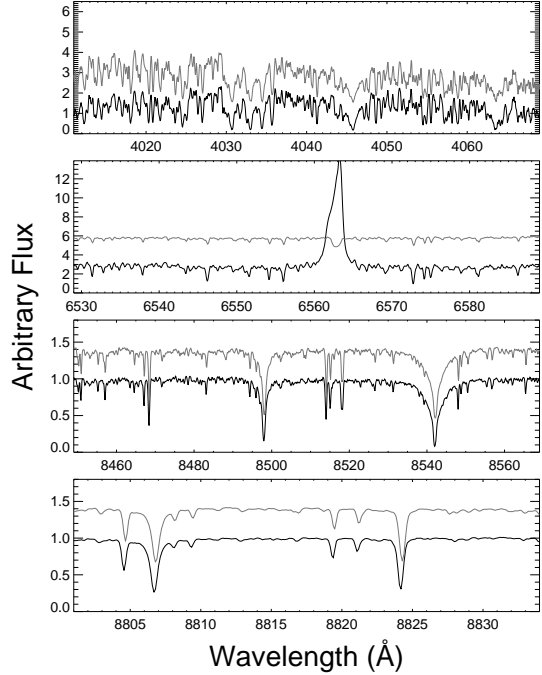}
\caption{Sections of the high-resolution APO echelle data are plotted in black. Overplotted in gray (with offset for clarity) is the spectrum of the M2 III spectral standard HD 148349.   \label{fig19}}
\end{figure}

The EG And APO  echelle data  cover the complete  wavelength region from 3500 \AA\ to 10400 \AA\ at high resolution. Since optical data diagnoses the photospheric layers of the giant, the extent to which the atmosphere of EG And giant differs to those of isolated giants can be viewed by a direct comparison of the spectrum with the spectrum taken of the spectral standard, HD 148349.  This was observed  on the same night with the same instrument. Displayed in Figure~\ref{fig19} are sections of the echelle data of EG And (black) and HD 148349 (gray and offset).

%\clearpage
%\clearpage

The similarity between both datasets is remarkable over the complete spectral range covered, including wavelength regions around molecular bands (top panel of Figure~\ref{fig18} for example), for weak lines from neutral species such as  \ion{Fe}{1} and \ion{Ti}{1} (examples in all panels) and in regions around  broadened features such as the \ion{Ca}{2} lines in the near-infrared spectral region (third panel from the top). Note that the EG And data was obtained at phase $\phi$=1.47, when the illuminated face of the giant is visible.

These \ion{Ca}{2} lines are sensitive to both the atmospheric pressure (surface gravity) and also to the metallicity. The only large  differences between the two spectra are in those regions where EG And displays strong nebular emission lines, such as illustrated in the wavelength region around the hydrogen Balmer line shown in the second panel of Figure~\ref{fig19}. This emission is due to the binary nature of EG And it is not a photospheric feature. Both stars, therefore, possess very similar photospheric parameters (including metallicity) and since both are high-velocity objects with similar abundances, they most likely belong to the same stellar population  \citep[both members of the old disk population; see][]{2001A&A...374..968M,1981Obs...101..172W}. 
HD 148349 is a bright (V magnitude$\sim$5.27), high-velocity (+99 km s$^{-1}$) M2 III spectral standard. \citet{1998NewA....3..137D} list the star as having a radius of  $R$=83 $R_{\sun}$, a T$_{eff}$=3,720 K, and a mass of $M$=2 $M_{\sun}$. From a direct viewing of the spectra, it appears that EG And giant possesses similar parameters. 

It is also worth noting that a rotational broadening of the narrow absorption lines in the EG And APO data imply a photospheric rotational velocity of $\sim$7.5 km s$^{-1}$. If the system is assumed to be tidally locked and co-rotating (the timescale for this process to occur is short for symbiotics and all can be assumed to co-rotate), then the implied  radius  of the giant is $\sim$70 $R_{\sun}$, consistent with that derived from the `eclipse' geometry \citep{1992A&A...260..156V}.

From this spectral comparison, it can be inferred that the atmospheric structure remains relatively unperturbed by the ionizing companion. It seems likely that any wind/chromospheric conditions derived from the ultraviolet data can be extrapolated to isolated stars as long as the effect of the dwarf radiation on the wind (originating beyond the photosphere) is understood.

\subsection{Effect of the Dwarf Radiation on the Giant Wind}

%In general terms we can claim to have a good understanding of  the significant spectral variations in the ultraviolet data, which can be explained in terms of the occultation of the hot component by the photosphere and extended atmosphere and wind of the giant component. However there remain 
There are variations which not related to orbital phase, such as the  behavior of the profiles of certain high-ionization features such as the the O VI resonance lines. In addition to these variations we also observe changes in the far-ultraviolet continuum fluxes which are not correlated to orbital phase. These variations were also observed with {\it{IUE}}, and are most notable in our final STIS spectrum  at phase $\phi=$4.50 where the ultraviolet continuum is observed to increase by a factor $\sim$2 above the normal level.  Although  EG And is relatively stable and has never been observed to undergo outburst,   we believe it is likely that these variations result from a variable accretion rate of giant wind material on to the dwarf. This would  result in variations in the the temperature and effective radius of the hot star and reproduce the observed continuum variations, as well as account for the time-variable broad high-ionization profiles. It is apparent that these processes are complicated to model and could have large effects on the cool giant wind. However, from our analysis of the P-Cygni profiles and the radial velocity  analysis of the unabsorbed broad lines, we can conclude that this material is located very close to the dwarf star. These features most likely diagnose dense, clumpy material accreting onto the white dwarf. It follows that this component can be readily separated from the material in the well-behaved outflow from the red giant in any in-depth analysis of the cool wind. 

The variable material being confined to a small region around the dwarf is also consistent with  the observed linewidths.  Assuming a white dwarf mass of 0.6 $M_{\sun}$ \citep{1992A&A...260..156V} and a gravitationally accelerated velocity of 300 km s$^{-1}$ (FWHM of broad  1640 \ion{He}{2} component), it emerges that the material must be closer than 3 $R_{\sun}$ from the dwarf's surface, which is less than 1$\%$ of the orbital separation. Using a velocity of 1000 km s$^{-1}$ (observed in higher ionization lines such as the \ion{O}{6} resonance profiles) the distance from the dwarf is further reduced to $\lesssim$ 1 $R_{\sun}$, or $\lesssim$0.003 of the binary separation. 

%To illustrate the point that this  material is located in a small volume relative to the system dimensions, we plot (Figure~\ref{fig19})  sections of two {\it{FUSE}} spectra taken at identical orbital phases ($\phi=$0.79) but at different orbital epochs. It is apparent that both spectra are are almost identical with the exceptions of the regions around the high-ionization, high-velocity lines. These features (resonance lines of \ion{Si}{4}, \ion{P}{5} and \ion{S}{4}) are observed to switch from P-Cygni to inverse P-Cygni form. Despite these drastic changes, the continuum and cool wind features are unaffected. This plot also demonstrates the repeatability of the observations of the red giant wind over several epochs. 

%\begin{figure}[!t]
%\figurenum{16}
%\epsscale{1.}
%\plotone{f20.eps}
%\caption{Sections of uneclipsed {\it{FUSE}} spectra of EG And taken at almost identical orbital phases (black - $\phi$=1.79; red - $\phi$=3.79), but two orbital epochs apart. The spectra are almost identical except for the high-ionization transitions. This is due to material close to the giant which undergoes variations unrelated to orbital phase. In general the material not located very close to the hot component remains stable over the different observed orbital cycles.  \label{fig20} }
%\end{figure} 

Yet another argument in favor of a relatively radiatively unperturbed wind lies in the observed ionization and excitation structure of the different lines of sight through the wind. The ionization level remains constant throughout the wind acceleration region and is symmetric about ultraviolet minimum. The fact that the dwarf is much less luminous than the giant (a factor of $\sim$60) would explain why the white dwarf photons do not ionize large regions of the cool wind. 

\subsection{Photoionization Modeling}
%\clearpage
\begin{figure}[!t]
\epsscale{1.}
\plotone{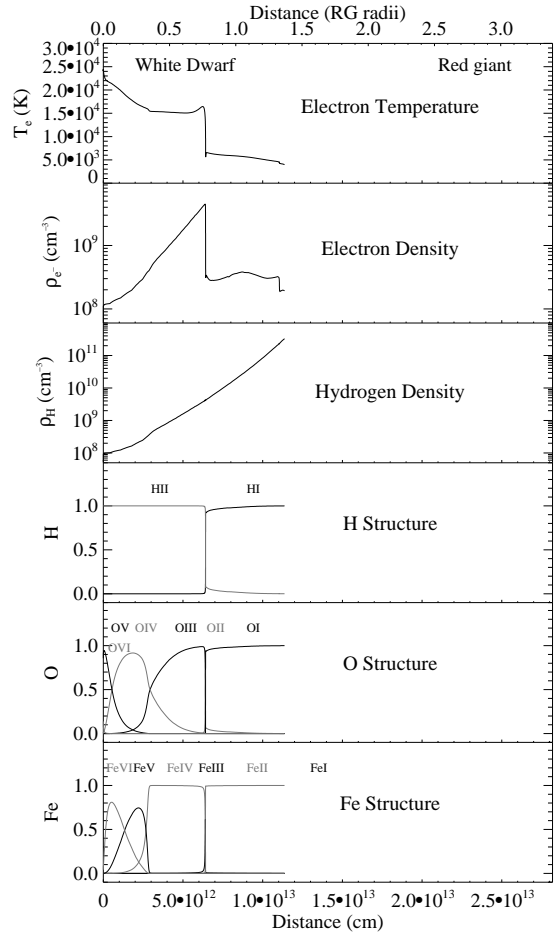}
\caption{Results of a CLOUDY photoionization modeling along the binary axis of both stars with dwarf on the left and and giant located on the right. The calculation is stopped when the electron temperature goes below 4,000K. The top three panels display the physical conditions in the affected part of the outer giant wind. The bottom three panels show the ionization structure of H, O and Fe.   \label{fig20}}
  \end{figure}
%\clearpage

\begin{figure}[!t]
\epsscale{1.}
\plotone{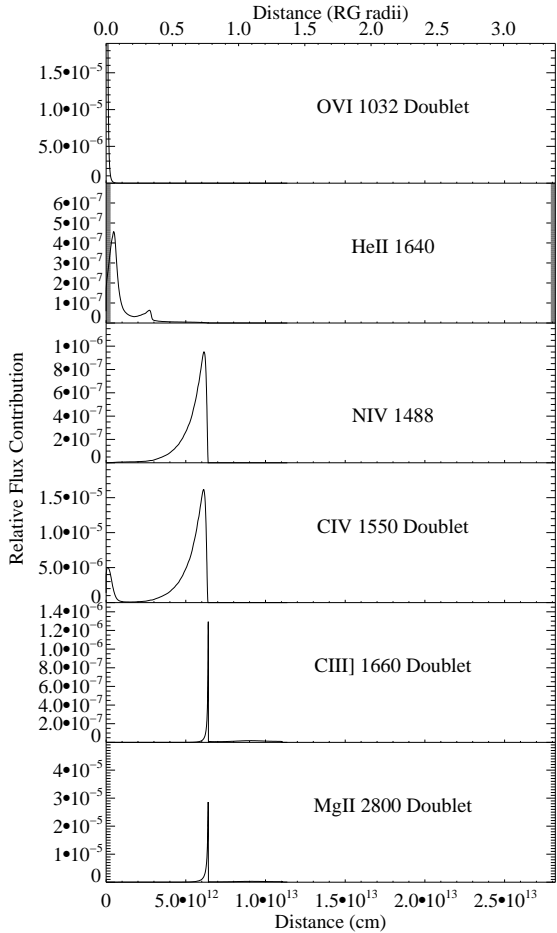}
\caption{Contribution functions of selected emission lines derived from the same photoionization model as described in Figure~\ref{fig20}. \label{fig21}}
\end{figure} 
%\clearpage

In order to analyze the radiative influence of the dwarf in the cool wind in a quantitative way, the CLOUDY \citep{1998PASP..110..761F} photoionization code was used to model the effects of the dwarf radiation on the wind. Using the stellar parameters described in Table 1 and a wind velocity law found for EG And by \citet{1991A&A...249..173V} (characterized by a steep acceleration $\sim$2.5 $R_{RG}$ above the photosphere and hereafter called the Vogel wind law) we have modeled the structure and conditions in the red giant wind along the binary axis between the two stars. Some of the findings are illustrated graphically in Figures~\ref{fig20} and \ref{fig21}.

The top three panels of Figure~\ref{fig20} show the effect of the ultraviolet radiation field on the conditions in the outer regions of the red giant wind. The dwarf is located at the left hand side of each panel while the surface of the giant is located on the right. It is noted that the wind is heated to a temperature above 20,000K close to the dwarf and drops as the radiation penetrates into denser material. Note that the hydrogen ionization boundary is located approximately 0.7  $R_{RG}$ from the dwarf's surface. It is at a distance of $\sim$1.4 $R_{RG}$ that the electron temperature (due solely to the dwarf photoionization) drops below 4,000K and the calculation is stopped. Beyond this point, the influence of the ultraviolet radiation is minimal. Although, the ionized region does not extend very far into the giant wind, it must be remembered that this calculation was carried out along the binary axis where the wind is most dense. For instance, the ultraviolet photons that escape in the opposite direction will ionize a much larger region due to the lower density of material. The lower three panels show the calculated ionization structure of hydrogen, oxygen and iron along the binary axis.%One can also see from Figure~\ref{fig20} the ionization structure of hydrogen, oxygen and iron along the binary axis. For these plots, the black, red, orange, magenta, green and blue lines correspond to different stages of ionization with black being neutral and blue corresponding to material that has lost five electrons. 

Presented in Figure~\ref{fig21} are the contribution functions for a number of emission lines along the binary axis. It can be noticed that the high ionization material associated with \ion{O}{6} and \ion{N}{5} resonance lines is located extremely close to the dwarf.  The \ion{He}{2} and \ion{C}{4} emission features also have a component located close to the dwarf, in addition to a component positioned further into the giant wind. This can explain the two-component profiles which are observed in the spectra for these transitions. For the majority of the nebular emission lines, most of the emitting material is located close to the hydrogen ionization boundary where the electron density is highest. This is due to a trade-off between the decreasing wind density as one moves further from the giant and the decreasing number of free electrons (due to photoionization) as one moves further from the dwarf. 

This photoionization model accounts for the relative emission line strengths observed in the data and, since it places the origin of most of the nebular emission lines at the regions of highest electron density which is along the binary axis, accounts for the variations in the line intensities around eclipse. The disappearance of the high-ionization features at eclipse is also explained by the placement of this gas very close to the dwarf surface.

\subsection{Mechanical Influence of Dwarf on the Wind}

One area in which the cool material is certainly affected by the presence of the dwarf, however, is through the redistribution of the wind material by the motion of the secondary. This redistribution is apparent in the asymmetry of the continuum fluxes and line strengths around eclipse, and whilst only a relatively slight effect, it must nevertheless be taken into account in any realistic wind models. In order to qualitatively understand the influence of the dwarf on the wind we examine the   hydrodynamical models of \citet{2000ASPC..204..331W}. The model which most closely describes the binary parameters of EG And \citep[low luminosity white dwarf model; see Figure 1 (left) of][]{2000ASPC..204..331W} predicts that he extent of the mechanical disturbance of the outflow is large. Due to a snow plough effect there is an increase in density of the material in front of the dwarf. There is a corresponding decrease in density in the wake of the dwarf, and at distances further from the dwarf, the outflow is greatly disturbed. However,  the material that is diagnosed by the phase-dependent absorption lines 
is located  {\em within} the binary orbit, where the wind is accelerated and is most dense. The outer, severely distorted parts of the outflow  are too tenuous to be viewed in absorption and, in any case, are beyond the point of initial wind acceleration and thus are of limited interest in terms of understanding the wind acceleration. It appears that, apart from the material directly in front of the dwarf, the wind within the binary orbit is relatively unaffected by the motion of the secondary component. This explains why the distribution of wind material as diagnosed by the data is not dramatically different between ingress and egress phases.

%\clearpage

\subsection{Conclusion}

We conclude that although it is possible to study cool winds through radio and ultraviolet observations of other binary systems, this dataset is unique in that it provides detailed thermal and dynamic information on the relatively unperturbed base of the wind where, crucially, the  wind acceleration processes occur. The ability to study the wind and chromosphere in both emission (near-ultraviolet low excitation emission) and absorption (through lines of sight) is also a major advantage. In addition to this, the observed absorption profiles are much cleaner and simpler to analyze than those obtained from studies of many other eclipsing binaries. In those cases the absorption profiles  are blended with the spectrum of the background secondary component and affected by the re-emission of scattered photons filling in the absorption profile, making a full 3-D radiative transfer analysis necessary \citep[][and references within]{1996ApJ...466..979B}. Due to the low luminosity of the EG And dwarf, the entirely different spectral characteristics of the stellar components and the relatively low excitation of the wind, the majority of the absorption lines for this dataset can be treated successfully with a pure absorption analysis. %A detailed analysis and modeling of the wind will be published in subsequent work.

\section{Summary}

We have reported on the first detailed ultraviolet and far-ultraviolet observations of EG And. This dataset is unique providing detailed thermal and dynamic information on the base of the wind where, crucially, the the wind acceleration processes occur. By choosing a symbiotic system with a low-luminosity dwarf many of the complications that typically hinder analysis of the lines of sight through giant winds are removed. The observational evidence (in the ultraviolet data) that the inner wind is relatively undisturbed by the dwarf is backed-up by analysis of optical data and photoionization modeling.

From analysis of the broad, high-ionization features in the data we discover  highly variable mass motions of relatively dense and ionized material. Indeed the broad absorption profiles of \ion{O}{6} and \ion{S}{6} are found to shift significantly between {\em FUSE} orbits ($\sim$90 minutes) in some cases. We find however, that  these processes are confined to a region very close to the hot component and that conditions in the giant's chromosphere and wind acceleration region are relatively unperturbed by the white dwarf radiation. By analyzing different lines of sight to the white dwarf through the base of the cool wind we see predominantly neutral (in hydrogen), moderately excited (lower absorption levels up to $\sim$5 eV) material. Although the ionization level is observed to remain constant throughout the base of the wind, there is evidence for an asymmetry in the distribution of material around the giant due to a mechanical redistribution by the motion of the dwarf.  The spatially resolved information provided by this dataset is useful for constraining mass-loss theory and understanding the processes at work in the chromospheres and wind acceleration regions of evolved stars. 

\acknowledgements

The authors wish to thank Gary Ferland, Henny Lamers and Peter Hauschildt for insightful discussions, as well as Brad Friel for advice on the  analysis of \ion{O}{6} profile variations. Also many thanks to referee Ulisse Munari for helpful comments and feedback.

This work is supported by Enterprise Ireland Basic Research grant SC/2002/370 from EU funded NDP.

{\em FUSE} data were obtained under the Guest Investigator Program and
supported
by NASA grants NAG5-8994 and NAG5-10403 to the Johns Hopkins University
(JHU).
The
NASA-CNES-CSA Far Ultraviolet Spectroscopic Explorer is operated for
NASA by the
JHU
under NASA contract NAS5-32985.  STIS data were obtained under the Guest
Observer program
STSI-GO0948701 provided by NASA from the Space
Telescope
Science
Institute, which is operated by the Association of Universities for
Research in
Astronomy, Inc., under NASA contract NAS 5-26555.

% FUSE data were obtained under the Guest Investigator Program (NASA grant ???) by the NASA-CNES-CSA FUSE mission, operated by Johns Hopkins University. 

\bibliographystyle{apj}

\bibliography{myref}

\end{document}

%% file: tab1.tex
\begin{deluxetable}{ccc}
\tabletypesize{\scriptsize}
%\rotate
\tablecaption{EG And Parameters\label{tbl-1}}
\tablewidth{0pt}
%\tablewidth{0pt}
\tablehead{
\colhead{Parameter} & \colhead{Value} & \colhead{References} 
}
\startdata
%\begin{table}[bht]
%  \caption{System parameters for EG And}
%  \label{tab:table}
%  \begin{center}
%    \leavevmode
%    \footnotesize
%    \begin{tabular}[h]{lr}
%      \hline \\[-5pt]
%      EG And Parameters      \\[+5pt]
%      \hline \\[-5pt]
    RG temperature & $\sim$ 3,700 K & 1 \\
    WD temperature & $\sim$ 75,000 K & 2 \\
RG Luminosity  & $\sim$ 950 $_{\sun}$ & 3 \\
WD Luminosity  & $\sim$ 16 $_{\sun}$ & 3 \\
Period & 482.6 days & 4 \\
Separation & $\sim$ 4.2 R$_{RG}$ & 3 \\
RG Radius & $\sim$ 75$_{\sun}$ & 3 \\
Inclination  & $>$ 70$^0$ & 3 \\
V$_{mag}$ & $\sim$ 7.1 & 5 \\
Orbit & circular & 6 \\
E(B-V) & 0.05 & 2 \\
%      \hline \\
%      \end{tabular}
%  \end{center}
%\end{table}
 \enddata    
\tablerefs{(1) \citet{2004AJ....128.2981K}; (2) \citet{1991A&A...248..458M}; (3) \citet{1992A&A...260..156V}; (4) \citet{2000AJ....119.1375F}; (5) \citet{2000A&AS..146..407B}; (6) \citet{1997MNRAS.291...54W}} 
 %     \hline \\
      \end{deluxetable}

%% file: tab2.tex
\begin{deluxetable}{lccr}
\tabletypesize{\scriptsize}
\tablecaption{EG And {\it{FUSE}} and STIS observations\label{tbl-2}}
\tablewidth{0pt}
\tablehead{
\colhead{Date} & \colhead{Telescope} & \colhead{Exp.\ time (Ks)} & \colhead{$\phi$ uv}
}
\startdata
2000 Jan 5 & FUSE & 11.0 & 1.79  \\
2000 Aug 6  & FUSE & 9.1 & 2.24  \\
2000 Nov 24 & FUSE & 5.5 &  2.47  \\
2001 Sep 3  & FUSE & 11.1 & 3.05  \\
2001 Sep 14 & FUSE & 7.5 & 3.07   \\
2001 Sep 28 & FUSE & 9.6 &3.10  \\
2001 Oct 23 & FUSE & 5.6 & 3.16  \\
2002 Aug 23 & FUSE & 10.9 & 3.79 \\
2002 Aug 28 & STIS & 4.6  & 3.80  \\
2002 Oct 16 & STIS & 4.6 &  3.90  \\
2002 Oct 20 & FUSE & 8.6 & 3.91  \\
2002 Oct 22 & FUSE & 6.7 & 3.91  \\
2002 Oct 23 & FUSE & 6.7 & 3.91  \\
2002 Dec 22 & STIS & 7.1 &  4.04  \\
2003 Jan 18 & STIS & 6.5 &  4.09 \\
2003 Jan 22 & FUSE & 3.4 & 4.10  \\
2003 Feb 6  & STIS & 7.6 &  4.13 \\
2003 Feb 16 & STIS & 7.3 & 4.15 \\
2003 Jul 31 & STIS & 4.5 &  4.50 \\
2003 Dec 01 & FUSE & 8.9 & 4.75  \\
\enddata
      
\tablecomments{Ephemeris from \citet{2000AJ....119.1375F}}
\end{deluxetable}

%% file: ms.bbl
\begin{thebibliography}{42}
\expandafter\ifx\csname natexlab\endcsname\relax\def\natexlab#1{#1}\fi

\bibitem[{{Baade}(1990)}]{1990iuea.rept...65B}
{Baade}, R. 1990, in Evolution in Astrophysics: IUE Astronomy in the Era of New
  Space Missions, ESA SP-310, 65--72

\bibitem[{{Baade} {et~al.}(1996){Baade}, {Kirsch}, {Reimers}, {Toussaint},
  {Bennett}, {Brown}, \& {Harper}}]{1996ApJ...466..979B}
{Baade}, R., {Kirsch}, T., {Reimers}, D., {Toussaint}, F., {Bennett}, P.~D.,
  {Brown}, A., \& {Harper}, G.~M. 1996, \apj, 466, 979

\bibitem[{{Belczy{\' n}ski} {et~al.}(2000){Belczy{\' n}ski}, {Miko{\l}ajewska},
  {Munari}, {Ivison}, \& {Friedjung}}]{2000A&AS..146..407B}
{Belczy{\' n}ski}, K., {Miko{\l}ajewska}, J., {Munari}, U., {Ivison}, R.~J., \&
  {Friedjung}, M. 2000, \aaps, 146, 407

\bibitem[{{Dixon} \& {Sahnow}(2003)}]{2003ASPC..295..241D}
{Dixon}, W.~V., \& {Sahnow}, D.~J. 2003, in ASP Conf. Ser. 295: Astronomical
  Data Analysis Software and Systems XII, 241

\bibitem[{{Dumm} \& {Schild}(1998)}]{1998NewA....3..137D}
{Dumm}, T., \& {Schild}, H. 1998, New Astronomy, 3, 137

\bibitem[{{Fekel} {et~al.}(2000){Fekel}, {Joyce}, {Hinkle}, \&
  {Skrutskie}}]{2000AJ....119.1375F}
{Fekel}, F.~C., {Joyce}, R.~R., {Hinkle}, K.~H., \& {Skrutskie}, M.~F. 2000,
  \aj, 119, 1375

\bibitem[{{Ferland} {et~al.}(1998){Ferland}, {Korista}, {Verner}, {Ferguson},
  {Kingdon}, \& {Verner}}]{1998PASP..110..761F}
{Ferland}, G.~J., {Korista}, K.~T., {Verner}, D.~A., {Ferguson}, J.~W.,
  {Kingdon}, J.~B., \& {Verner}, E.~M. 1998, \pasp, 110, 761

\bibitem[{{Judge} \& {Jordan}(1991)}]{1991ApJS...77...75J}
{Judge}, P.~G., \& {Jordan}, C. 1991, \apjs, 77, 75

\bibitem[{{Kenyon}(1986)}]{1986syst.book.....K}
{Kenyon}, S.~J. 1986, {The symbiotic stars} (Cambridge and New York, Cambridge
  University Press, 1986, 295)

\bibitem[{{Kenyon} \& {Fernandez-Castro}(1987)}]{1987AJ.....93..938K}
{Kenyon}, S.~J., \& {Fernandez-Castro}, T. 1987, \aj, 93, 938

\bibitem[{{Kenyon} {et~al.}(1986){Kenyon}, {Fernandez-Castro}, \&
  {Stencel}}]{1986AJ.....92.1118K}
{Kenyon}, S.~J., {Fernandez-Castro}, T., \& {Stencel}, R.~E. 1986, \aj, 92,
  1118

\bibitem[{{Keyes} \& {Preblich}(2004)}]{2004AJ....128.2981K}
{Keyes}, C.~D., \& {Preblich}, B. 2004, \aj, 128, 2981

\bibitem[{{Kurucz} \& {Bell}(1995)}]{1995all..book.....K}
{Kurucz}, R.~L., \& {Bell}, B. 1995, {Atomic line list} (Kurucz CD-ROM no.\ 23,
  Cambridge, MA: Smithsonian Astrophysical Observatory)

\bibitem[{{Mennessier} {et~al.}(2001){Mennessier}, {Mowlavi}, {Alvarez}, \&
  {Luri}}]{2001A&A...374..968M}
{Mennessier}, M.~O., {Mowlavi}, N., {Alvarez}, R., \& {Luri}, X. 2001, \aap,
  374, 968

\bibitem[{{Mikolajewska} {et~al.}(1988){Mikolajewska}, {Friedjung}, {Kenyon},
  \& {Viotti}}]{1988syph.book.....M}
{Mikolajewska}, J., {Friedjung}, M., {Kenyon}, S.~J., \& {Viotti}, R. 1988,
  {The symbiotic phenomenon} (ASSL Vol.~145: IAU Colloq.~103: The Symbiotic
  Phenomenon)

\bibitem[{{Moos} {et~al.}(2000){Moos}, {Cash}, {Cowie}, {Davidsen}, {Dupree},
  {Feldman}, {Friedman}, {Green}, {Green}, {Gry}, {Hutchings}, {Jenkins},
  {Linsky}, {Malina}, {Michalitsianos}, {Savage}, {Shull}, {Siegmund}, {Snow},
  {Sonneborn}, {Vidal-Madjar}, {Willis}, {Woodgate}, {York}, {Ake},
  {Andersson}, {Andrews}, {Barkhouser}, {Bianchi}, {Blair}, {Brownsberger},
  {Cha}, {Chayer}, {Conard}, {Fullerton}, {Gaines}, {Grange}, {Gummin},
  {Hebrard}, {Kriss}, {Kruk}, {Mark}, {McCarthy}, {Morbey}, {Murowinski},
  {Murphy}, {Oegerle}, {Ohl}, {Oliveira}, {Osterman}, {Sahnow}, {Saisse},
  {Sembach}, {Weaver}, {Welsh}, {Wilkinson}, \& {Zheng}}]{2000ApJ...538L...1M}
{Moos}, H.~W., {Cash}, W.~C., {Cowie}, L.~L., {Davidsen}, A.~F., {Dupree},
  A.~K., {Feldman}, P.~D., {Friedman}, S.~D., {Green}, J.~C., {Green}, R.~F.,
  {Gry}, C., {Hutchings}, J.~B., {Jenkins}, E.~B., {Linsky}, J.~L., {Malina},
  R.~F., {Michalitsianos}, A.~G., {Savage}, B.~D., {Shull}, J.~M., {Siegmund},
  O.~H.~W., {Snow}, T.~P., {Sonneborn}, G., {Vidal-Madjar}, A., {Willis},
  A.~J., {Woodgate}, B.~E., {York}, D.~G., {Ake}, T.~B., {Andersson}, B.-G.,
  {Andrews}, J.~P., {Barkhouser}, R.~H., {Bianchi}, L., {Blair}, W.~P.,
  {Brownsberger}, K.~R., {Cha}, A.~N., {Chayer}, P., {Conard}, S.~J.,
  {Fullerton}, A.~W., {Gaines}, G.~A., {Grange}, R., {Gummin}, M.~A.,
  {Hebrard}, G., {Kriss}, G.~A., {Kruk}, J.~W., {Mark}, D., {McCarthy}, D.~K.,
  {Morbey}, C.~L., {Murowinski}, R., {Murphy}, E.~M., {Oegerle}, W.~R., {Ohl},
  R.~G., {Oliveira}, C., {Osterman}, S.~N., {Sahnow}, D.~J., {Saisse}, M.,
  {Sembach}, K.~R., {Weaver}, H.~A., {Welsh}, B.~Y., {Wilkinson}, E., \&
  {Zheng}, W. 2000, \apjl, 538, L1

\bibitem[{{Morton}(2003)}]{2003ApJS..149..205M}
{Morton}, D.~C. 2003, \apjs, 149, 205

\bibitem[{{Muerset} {et~al.}(1991){Muerset}, {Nussbaumer}, {Schmid}, \&
  {Vogel}}]{1991A&A...248..458M}
{Muerset}, U., {Nussbaumer}, H., {Schmid}, H.~M., \& {Vogel}, M. 1991, \aap,
  248, 458

\bibitem[{{Munari}(1989)}]{1989A&A...208...63M}
{Munari}, U. 1989, \aap, 208, 63

\bibitem[{{Munari}(1993)}]{1993A&A...273..425M}
---. 1993, \aap, 273, 425

\bibitem[{{Nussbaumer} {et~al.}(1995){Nussbaumer}, {Schmutz}, \&
  {Vogel}}]{1995A&A...293L..13N}
{Nussbaumer}, H., {Schmutz}, W., \& {Vogel}, M. 1995, \aap, 293, L13

\bibitem[{{Oliversen} {et~al.}(1985){Oliversen}, {Anderson}, {Slovak}, \&
  {Stencel}}]{1985ApJ...295..620O}
{Oliversen}, N.~A., {Anderson}, C.~M., {Slovak}, M.~H., \& {Stencel}, R.~E.
  1985, \apj, 295, 620

\bibitem[{{Raassen} \& {Uylings}(1998)}]{1998yCat..33400300R}
{Raassen}, A.~J.~J., \& {Uylings}, P.~H.~M. 1998, VizieR Online Data Catalog,
  334, 300

\bibitem[{{Reimers}(1987)}]{1987IAUS..122..307R}
{Reimers}, D. 1987, in IAU Symp.\ 122: Circumstellar Matter, ed.\ I.\
  Appenzeller, \& C.\ Jordan (Dordrecht: Kluwert), 307--318

\bibitem[{{Sahnow} {et~al.}(2000){Sahnow}, {Moos}, {Ake}, {Andersen},
  {Andersson}, {Andre}, {Artis}, {Berman}, {Blair}, {Brownsberger}, {Calvani},
  {Chayer}, {Conard}, {Feldman}, {Friedman}, {Fullerton}, {Gaines}, {Gawne},
  {Green}, {Gummin}, {Jennings}, {Joyce}, {Kaiser}, {Kruk}, {Lindler}, {Massa},
  {Murphy}, {Oegerle}, {Ohl}, {Roberts}, {Romelfanger}, {Roth}, {Sankrit},
  {Sembach}, {Shelton}, {Siegmund}, {Silva}, {Sonneborn}, {Vaclavik}, {Weaver},
  \& {Wilkinson}}]{2000ApJ...538L...7S}
{Sahnow}, D.~J., {Moos}, H.~W., {Ake}, T.~B., {Andersen}, J., {Andersson},
  B.-G., {Andre}, M., {Artis}, D., {Berman}, A.~F., {Blair}, W.~P.,
  {Brownsberger}, K.~R., {Calvani}, H.~M., {Chayer}, P., {Conard}, S.~J.,
  {Feldman}, P.~D., {Friedman}, S.~D., {Fullerton}, A.~W., {Gaines}, G.~A.,
  {Gawne}, W.~C., {Green}, J.~C., {Gummin}, M.~A., {Jennings}, T.~B., {Joyce},
  J.~B., {Kaiser}, M.~E., {Kruk}, J.~W., {Lindler}, D.~J., {Massa}, D.,
  {Murphy}, E.~M., {Oegerle}, W.~R., {Ohl}, R.~G., {Roberts}, B.~A.,
  {Romelfanger}, M.~L., {Roth}, K.~C., {Sankrit}, R., {Sembach}, K.~R.,
  {Shelton}, R.~L., {Siegmund}, O.~H.~W., {Silva}, C.~J., {Sonneborn}, G.,
  {Vaclavik}, S.~R., {Weaver}, H.~A., \& {Wilkinson}, E. 2000, \apjl, 538, L7

\bibitem[{{Schmid}(2003)}]{2003ASPC..303..343S}
{Schmid}, H.~M. 2003, in Astronomical Society of the Pacific Conference Series,
  `Symbiotic Stars Probing Stellar Evolution', edited by R. L. M. Corradi, R.
  Mikolajewska and T. J. Mahoney, ISBN: 1-58381-152-4, 343

\bibitem[{{Seaquist} \& {Taylor}(1990)}]{1990ApJ...349..313S}
{Seaquist}, E.~R., \& {Taylor}, A.~R. 1990, \apj, 349, 313

\bibitem[{{Sion} \& {Ready}(1992)}]{1992PASP..104...87S}
{Sion}, E.~M., \& {Ready}, C.~J. 1992, \pasp, 104, 87

\bibitem[{{Skopal}(2005)}]{2005A&A...440..995S}
{Skopal}, A. 2005, \aap, 440, 995

\bibitem[{{Skopal} {et~al.}(1991){Skopal}, {Chochol}, {Vittone}, {Blanco}, \&
  {Mammano}}]{1991A&A...245..531S}
{Skopal}, A., {Chochol}, D., {Vittone}, A.~A., {Blanco}, C., \& {Mammano}, A.
  1991, \aap, 245, 531

\bibitem[{{Smith}(1980)}]{1980ApJ...237..831S}
{Smith}, S.~E. 1980, \apj, 237, 831

\bibitem[{{Stencel}(1984)}]{1984ApJ...281L..75S}
{Stencel}, R.~E. 1984, \apjl, 281, L75

\bibitem[{{Stencel} \& {Sahade}(1980)}]{1980ApJ...238..929S}
{Stencel}, R.~E., \& {Sahade}, J. 1980, \apj, 238, 929

\bibitem[{{Tomov}(1995)}]{1995MNRAS.272..189T}
{Tomov}, N.~A. 1995, \mnras, 272, 189

\bibitem[{{van Buren} {et~al.}(1994){van Buren}, {Dgani}, \&
  {Noriega-Crespo}}]{1994AJ....108.1112V}
{van Buren}, D., {Dgani}, R., \& {Noriega-Crespo}, A. 1994, \aj, 108, 1112

\bibitem[{{Vogel}(1991)}]{1991A&A...249..173V}
{Vogel}, M. 1991, \aap, 249, 173

\bibitem[{{Vogel}(1993)}]{1993A&A...274L..21V}
---. 1993, \aap, 274, L21

\bibitem[{{Vogel} {et~al.}(1992){Vogel}, {Nussbaumer}, \&
  {Monier}}]{1992A&A...260..156V}
{Vogel}, M., {Nussbaumer}, H., \& {Monier}, R. 1992, \aap, 260, 156

\bibitem[{{Walder} \& {Folini}(2000)}]{2000ASPC..204..331W}
{Walder}, R., \& {Folini}, D. 2000, in ASP Conf. Ser. 204: Thermal and
  Ionization Aspects of Flows from Hot Stars, Ed.\ by H.\ Lamers \& A.\ Sapar,
  331

\bibitem[{{Wallerstein}(1981)}]{1981Obs...101..172W}
{Wallerstein}, G. 1981, The Observatory, 101, 172

\bibitem[{{Wilson} \& {Vaccaro}(1997)}]{1997MNRAS.291...54W}
{Wilson}, R.~E., \& {Vaccaro}, T.~R. 1997, \mnras, 291, 54

\bibitem[{{York}(1995)}]{1995AAS...186.4404Y}
{York}, D.~G. 1995, Bulletin of the American Astronomical Society, 27, 874

\end{thebibliography}
